\documentclass[10pt,journal,compsoc]{IEEEtran}
\usepackage[noadjust]{cite}

\ifCLASSINFOpdf
\else
\fi
\usepackage{multirow}
\usepackage[table,xcdraw]{xcolor}
\usepackage{listings}

\definecolor{codegreen}{rgb}{0,0.6,0}
\definecolor{codegray}{rgb}{0.5,0.5,0.5}
\definecolor{codepurple}{rgb}{0.58,0,0.82}
\definecolor{backcolour}{rgb}{0.95,0.95,0.92}

\lstdefinestyle{mystyle}{
    backgroundcolor=\color{backcolour},   
    commentstyle=\color{codegreen},
    keywordstyle=\color{magenta},
    numberstyle=\tiny\color{codegray},
    stringstyle=\color{codepurple},
    basicstyle=\ttfamily\footnotesize,
    breakatwhitespace=false,         
    breaklines=true,                 
    captionpos=b,                    
    keepspaces=true,                 
    numbers=left,                    
    numbersep=5pt,                  
    showspaces=false,                
    showstringspaces=false,
    showtabs=false,                  
    tabsize=2
}

\usepackage{graphicx}

\begin{document}
\bstctlcite{IEEEexample:BSTcontrol}

\newcommand{\apachebenchname}{{ApacheCryptoAPI-Bench}}

\newcommand*\circleinsec[1]{
  \tikz[baseline=(X.base)] 
  \node (X) [draw, shape=circle, inner sep=0pt, fill=red, text=black] {\strut #1};
}
%
\title{Evaluation of Static Vulnerability Detection Tools with Java Cryptographic API Benchmarks}

%
%
%
%

\author{Sharmin~Afrose,
        ~Ya~Xiao,
        ~Sazzadur~Rahaman, 
        ~Barton P. Miller,
        ~Danfeng~(Daphne)~Yao
\IEEEcompsocitemizethanks{\IEEEcompsocthanksitem S. Afrose, Y. Xiao and D. Yao are with the Department
of Computer Science, Virginia Tech, Blacksburg,
VA, 24060.\protect\\
E-mail: sharminafrose@vt.edu, yax99@vt.edu, danfeng@vt.edu
\IEEEcompsocthanksitem S. Rahaman is with the Department of Computer Science, University of Arizona, Tucson, AZ, 85721.\protect\\
E-mail: sazz@cs.arizona.edu
\IEEEcompsocthanksitem  B. P. Miller is with the Computer Sciences Department, University of Wisconsin-Madison, Madison, WI, 53706.\protect\\
E-mail: bart@cs.wisc.edu
}

}

\IEEEtitleabstractindextext{%
\begin{abstract}
Several studies showed that misuses of cryptographic APIs are common in real-world code (e.g., Apache projects and Android apps). There exist several open-sourced and commercial security tools that automatically screen Java programs to detect misuses. To compare their accuracy and security guarantees, we develop two comprehensive benchmarks named CryptoAPI-Bench and \apachebenchname{}. CryptoAPI-Bench consists of 181 unit test cases that cover basic cases, as well as complex cases, including interprocedural, field sensitive, multiple class test cases, and path sensitive data flow of misuse cases. The benchmark also includes correct cases for testing false-positive rates. The \apachebenchname{} consists of 121 cryptographic cases from 10 Apache projects. We evaluate four tools, namely, SpotBugs, CryptoGuard, CrySL, and Coverity using both benchmarks. We present their performance and comparative analysis.  The \apachebenchname{} also examines the scalability of the tools. Our benchmarks are useful for advancing state-of-the-art solutions in the space of misuse detection.
\end{abstract}

}

\maketitle

\IEEEdisplaynontitleabstractindextext

%
\IEEEpeerreviewmaketitle

\vspace{-5mm}
\IEEEraisesectionheading{\section{Introduction}\label{sec:introduction}}

%
%
%
%

Various studies have shown that a vast majority of Java and Android applications misuse cryptographic libraries and APIs, causing devastating security and privacy implications. The most pervasive cryptographic misuses include exposed secrets (e.g., secret keys and passwords), predictable random numbers, use of insecure crypto primitives, vulnerable certificate verification~\cite{DBLP:conf/ccs/FahlHMSBF12, DBLP:conf/ccs/GeorgievIJABS12, DBLP:conf/ccs/EgeleBFK13, Meng-ICSE-2018, DBLP:journals/corr/abs-1806-06881, DBLP:conf/secdev/RahamanY17}.

Several studies showed that the prominent causes for cryptographic misuses are the deficiency in understanding of security API usage~\cite{SP-crypto-API-2017, Meng-ICSE-2018}, complex API designs~\cite{SP-crypto-API-2017, Bodden-crypto-API-2015}, the lack of cybersecurity training~\cite{Meng-ICSE-2018}, insecure code generation tools~\cite{DBLP:conf/sp/OltroggeDSAFRPB18} and insecure/misleading suggestions in Stack Overflow~\cite{DBLP:conf/sp/AcarBFKMS16,Meng-ICSE-2018}. The reality is that most developers, with tight project deadlines and short product turnaround time, spend little effort on improving their knowledge or hardening their code for long-term benefits~\cite{DBLP:conf/soups/AssalC18}.
Recognizing these practical barriers, automatic cryptographic code generation~\cite{DBLP:conf/kbse/KrugerNRAMBGGWD17}, and misuse detection tools~\cite{DBLP:journals/corr/abs-1806-06881} play a significant role in assisting developers with writing and maintaining secure code.

The security community has produced several impressive static (e.g., CryptoLint~\cite{DBLP:conf/ccs/EgeleBFK13}, CrySL~\cite{DBLP:conf/ecoop/KrugerS0BM18}, FixDroid~\cite{DBLP:conf/ccs/NguyenWABWF17}, MalloDroid~\cite{DBLP:conf/ccs/FahlHMSBF12}, CryptoGuard~\cite{DBLP:journals/corr/abs-1806-06881}) and dynamic code screening tools (e.g., Crylogger~\cite{piccolboni2020crylogger}, SMV-Hunter~\cite{DBLP:conf/ndss/SounthirarajSGLK14}, and AndroSSL~\cite{DBLP:conf/fps/GagnonFFDOB15}) to detect API misuses in Java. The static analysis does not require a program to execute, rather it is performed on a version of the code (e.g., source code, intermediate representations or binary). Many abstract security rules are reducible to concrete program properties that are enforceable via generic static analysis techniques~\cite{DBLP:conf/sp/AshcraftE02, DBLP:journals/corr/abs-1806-06881}. Consequently, static analysis tools have the potential to cover a wide range of security rules. In contrast, dynamic analysis tools require one to execute a program and spend a significant effort to trigger and detect specific misuse symptoms at runtime. Hence, dynamic analysis tools may be limited in their coverage. A code screening tool needs to be scalable with wide coverage. Thus, static analysis-based tools are usually more favorable than their dynamic counterparts. 

\vspace{-0.5mm}
However, a major weakness of static analysis tools is their tendency to produce false alerts. False alerts substantially diminished the value of a tool. To reduce the number of false positives, most of the static analysis tools offer a trade-off between completeness and scalability~\cite{DBLP:conf/uss/MachirySCSKV17}. We define \textit{completeness} as the ability to detect all the misuse instances and \textit{scalability} as the ability to induce low computational overhead to analyze large code-bases. Designing tools that would produce fewer false positives and false negatives with smaller computational overhead help the real-world deployment.

To advance and monitor the scientific progress of domains to produce effective tools, a mechanism for comparative studies is required. Unfortunately, for automatic detection of cryptographic API misuses, no suitable mechanism or benchmark exists. Such a benchmark needs to have several requirements:
{\em i)} It should cover a wide range of misuse instances. {\em ii)} It should cover interesting program properties (e.g., flow-, context-, field-, path-sensitivity, etc.)~\cite{wiki:dataflow, 10.1145/3290361}. These are different detection capabilities required for capturing certain vulnerabilities. {\em iii)} Test cases should be written in easily compilable source codes, so that both source code and binary code analysis tools can be easily evaluated. 

None of the existing benchmarks follows these criteria (e.g., DroidBench~\cite{DBLP:conf/pldi/ArztRFBBKTOM14}, Ghera~\cite{DBLP:conf/promise/MitraR17}). For example, DroidBench~\cite{DBLP:conf/pldi/ArztRFBBKTOM14} only contains binaries. Ghera~\cite{DBLP:conf/promise/MitraR17} has sources of provided Android apps. However, both DroidBench and Ghera barely cover cryptographic API misuses.

In this paper, we present two benchmarks for cryptographic API misuses. The first one is CryptoAPI-Bench, a comprehensive benchmark for comparing the quality of cryptographic vulnerability detection tools. It consists of 181 unit test cases covering 18 types of cryptographic misuses. Several test cases include interesting program properties~\cite{wiki:dataflow, 10.1145/3290361}. Flow-sensitive correctly computes and analyzes the order of statements in a program. Path-sensitivity analysis computes different dataflow analysis information dependent on conditional branch statements. Field-sensitive analysis distinguishs two fields containing the same object in a class. A context-sensitive analysis is interprocedural analysis that analyzes the target of a function call. 

The second one is \apachebenchname{} which is built upon 10 real-world Apache projects. It contains early versions of activemq-artemis, deltaspike, directory-server, manifoldcf, meecrowave, spark, tika, tomee, wicket projects. We identify 121 crypto cases in them, including 79 basic cases and 42 advanced cases. 

We run CryptoAPI-Bench and \apachebenchname{} on four static analysis tools (i.e., SpotBugs~\cite{spotbugs}, CryptoGuard, CrySL, and Coverity~\cite{coverity}) and perform a comparative analysis of these tools. These tools are {\em i)} capable of detecting cryptographic misuse vulnerabilities and {\em ii)} open-sourced and/or provide free evaluation license. CrySL and CryptoGuard are open-sourced research prototypes that are actively being maintained to improve their accuracy and coverage. SpotBugs is also an actively maintained open-source project, which is the successor of FindBugs. Coverity is one of the most popular static analysis platforms for decades. 

Our main technical contributions are summarized as follows.
\vspace{-1mm}
\begin{itemize}
    \item We provide a benchmark named CryptoAPI-Bench, which consists of 181 test cases covering 16 types of Cryptographic and SSL/TLS API misuse vulnerabilities. CryptoAPI-Bench utilized various interesting program properties (e.g., field-, context-, and path-sensitivity) to produce a diverse set of test cases. Our benchmark is open-sourced and can be found on GitHub~\cite{cryptoapibench-github}.
    
    \item We provide another benchmark named \apachebenchname{} for checking the scalability property of the cryptographic vulnerability detection tools. We document 121 test cases covering 12 types of Cryptographic and SSL/TLS API misuse vulnerabilities from 10 real-world Apache projects. The detailed information regarding \apachebenchname{} can be found on GitHub~\cite{apachecryptoapibench-github}.
    
    \item We evaluate four static analysis tools that are capable of detecting cryptographic misuse vulnerabilities. Our experimental evaluation revealed some interesting insights. For complex cases, specialized tools (e.g., CryptoGuard, CrySL) detect more cryptographic misuses and cover more rules than general-purpose tools (e.g., SpotBugs, Coverity). Currently, none of these tools supports path-sensitive analysis.

\end{itemize}

\vspace{-1mm}

A preliminary version of the work appeared in the Proceedings of the 2019 ACM Conference on Computer and Communications Security (CCS)~\cite{DBLP:journals/corr/abs-1806-06881} and 2019 IEEE Secure Development Conference (SecDev)~\cite{afrose2019cryptoapi}. We expanded the conference version by adding a new benchmark ApacheCryptoAPI-Bench (Section 4, Table 2) that contains complex real-world Java programs and we test four static tools' performance in real-world code (Section 6.4, Table 7, Table 8). For CryptoAPI-Bench, We also add two new misuse categories (Section 2.17, Section 2.18), 11 new test cases (Table 1), and update tools' performance evaluation (Table 4, Table 5, Table 6).

The remainder of this paper is organized as follows. Section~\ref{sec:threat_model} describes cryptographic API misuse categories. Section~\ref{sec:design} and Section~\ref{sec:design_apache} outlines the design of CryptoAPI-Bench and \apachebenchname{}. Section~\ref{sec:existing_tools} reviews existing cryptographic vulnerability detection tools. Section~\ref{sec:evaluation_analysis} presents the evaluation and performance analysis of the tools on the benchmarks. Discussion is given in Section~\ref{sec:discussion}. Section~\ref{sec:related_work} describes the related works.  Finally, Section~\ref{conclusion} concludes this paper.

\vspace{-4mm}
\section{Crypto API misuse categories}
\label{sec:threat_model}
\vspace{-1mm}
In this section, we discuss 18 Java cryptographic API misuse categories. We got the insights of these misuse categories from previous literature~\cite{DBLP:journals/corr/abs-1806-06881, DBLP:conf/ecoop/KrugerS0BM18, DBLP:conf/ccs/NguyenWABWF17}, NIST documents~\cite{nistdoc,244246,nistcsrc}, and other blogs~\cite{sonarsource}. We describe reasons for vulnerability and possible secure solutions for these misuse categories. 

\vspace{1mm}
\textbf{\noindent{\em 2.1 Cryptographic Keys:}}
For encryption, it is expected to use an unpredictable key using \texttt{javax.crypto.spec.SecretKeySpec} API that takes a byte array as input. If the Byte array is constant or hardcoded inside the code, the adversary can easily read the cryptographic key and may obtain sensitive information. Therefore, an unpredictable byte array should be used as a parameter in SecretKeySpec to generate a secure key.  

\vspace{1mm}
\textbf{\noindent{\em 2.2 Passwords in Password-based Encryption:}}
Password-based Encryption (PBE) is a popular technique of generating a strong secret key using \texttt{javax.crypto.spec.PBEKeySpec} API. It takes three parameters (i.e., password, salt, and iteration count). However, if a hardcoded or constant password is used in the code, then malicious attackers may obtain the password and predict the key~\cite{DBLP:conf/ccs/EgeleBFK13}. Therefore, an unpredictable password should be used as a parameter in PBEKeySpec. 


\vspace{1mm}
\textbf{\noindent{\em 2.3 Passwords in KeyStore:}}
Cryptographic keys and certificates are sometimes stored using \texttt{java.security.KeyStore} API. The KeyStore employs a password to get access to the stored keys and certificates. However, if a hardcoded or constant password is used for KeyStore in the code, it poses a security threat of revealing keys and certificates stored in the KeyStore. Therefore, an unpredictable random password should be used in KeyStore.

\vspace{1mm}
\textbf{\noindent{\em 2.4 Hostname Verifier:}}
HostnameVerifier in \texttt{javax.net.ssl.HostnameVerifier} API verifies the hostname by checking the hostname's authentication and identification. In some cases, verify() method of HostnameVerifier class is set to return true by default so that the verification method can quickly get past an exception. However, this arrangement causes a security threat, where URL spoofing~\cite{UrlSpoofing} attacks can be possible. URL spoofing makes it simpler for numerous cyber-attacks (e.g., identity theft, phishing). 

    

\textbf{\noindent{\em 2.5 Certificate Validation:}}
Empty methods are often implemented in \texttt{javax.net.ssl.X509TrustManager} interface to connect quickly and easily with clients and remote servers without any certificate validation. In that case, the TrustManager accepts and trusts every entity including the entity that is not signed by a trusted certificate authority. It may cause Man-in-the-middle (MitM) attacks~\cite{BugsPatternTrustManager,DBLP:conf/ccs/FahlHMSBF12}.

\vspace{1mm}
\textbf{\noindent{\em 2.6 SSL Sockets:}}
\texttt{javax.net.ssl.SSLSocket} connects a specific host to a specific port. However, before the connection, the hostname of the server should be verified and authenticated using \texttt{javax.net.ssl.HostnameVerifier} API. However, incorrect implementation omits the hostname verification when the socket is created~\cite{HostnameVerificationSSLSocket,DBLP:conf/ccs/GeorgievIJABS12}. 

\vspace{1mm}
\textbf{\noindent{\em 2.7 Hypertext Transfer Protocol:}}
HyperText Transfer Protocol (HTTP) sends a request to a server to retrieve a web page. However, HTTP allows hackers to intercept and read sensitive information~\cite{barton2006transferring}. Therefore, it is recommended to use HyperText Transfer Protocol Secure (HTTPS) that utilizes a secured socket layer to encrypt sensitive information. 


\vspace{1mm}
\textbf{\noindent{\em 2.8 Pseudorandom Number Generator (PRNG):}}
The generation of a pseudorandom number using \texttt{java.util.Random} is vulnerable as the generated random number is not completely random, because it uses a definite mathematical algorithm (Knuth's subtractive random number generator algorithm~\cite{knuth2014art}) that is proven to be insecure. To solve the problem, \texttt{java.security.SecureRandom} provides non-deterministic and unpredictable random numbers. 


\vspace{1mm}
\textbf{\noindent{\em 2.9 Seeds in Pseudorandom Number Generator (PRNG)}}
While using \texttt{java.security.SecureRandom}, if a constant or static seed is provided in SecureRandom, then it is possible to have the same outcome on every run. Therefore, developers should use a non-deterministic random seed. 

\vspace{1mm}
\textbf{\noindent{\em 2.10 Salts in Password-based encryption (PBE):}}
\texttt{javax.crypto.spec.PBEParameterSpec} API takes salt as one of the parameters for Password-based encryption. Using constant or static salts increases the possibility of a dictionary attack. The salt should be a random number that produces a random and unpredictable key. 


\vspace{1mm}
\textbf{\noindent{\em 2.11 Mode of Operation:}}
The Electronic Codebook (ECB) mode of operation is insecure to use in \texttt{javax.crypto.Cipher} as ECB-encrypted ciphertext can leak information about the plaintext. Instead of ECB, Cipher Block Chaining (CBC) or Galois/Counter Mode (GCM) is more secure to use.

\vspace{1mm}
\textbf{\noindent{\em 2.12 Initialization Vector (IV):}}
The initialization vector (IV) is used during encryption and decryption with several modes of operation. Static/constant initialization vector introduces vulnerabilities for CBC mode of operation. Therefore, it is suggested to use an unpredictable random initialization vector in \texttt{crypto.spec.IvParameterSpec API}. Note that, for several modes of operation (e.g., CTR, CBC-MAC), unpredictable random IV is not required.     


\begin{table*}[!h]

\caption{CryptoAPI-Bench: Summary of unit test cases. There are 181 unit test cases with 45 basic cases and 136 advanced cases (interprocedural, field sensitive, combined case, path sensitive, miscellaneous, and multiple class test cases). Total test cases per group and misuse categories are summarized here. Details information are presented in Section~\ref{sec:design}. }
\vspace{-3mm}
\label{table:rule_vs_group}
\centering
\begin{tabular}{|c|l|c|c|c|c|c|c|c|c|c|}
\hline \rowcolor{lightgray}
No. & \textbf{Misuse Categories}               & \textbf{\begin{tabular}[c]{@{}c@{}}Basic \\ Cases\end{tabular}} & \textbf{\begin{tabular}[c]{@{}c@{}}Two-\\ Interproc.\end{tabular}} & \textbf{\begin{tabular}[c]{@{}c@{}}Three-\\ Interproc.\end{tabular}} & \textbf{\begin{tabular}[c]{@{}c@{}}Field \\ Sensitive\end{tabular}} & \textbf{\begin{tabular}[c]{@{}c@{}}Combined \\ Case\end{tabular}} & \textbf{\begin{tabular}[c]{@{}c@{}}Path \\ Sensitive\end{tabular}} & \textbf{Misc.} & \textbf{\begin{tabular}[c]{@{}c@{}}Multiple \\ Class\end{tabular}} & \textbf{\begin{tabular}[c]{@{}c@{}}Total  Cases \\ per Categories\end{tabular}} \\ \hline 
1   & Cryptographic Key                 & 2                                                               & 1                                                                       & 1                                                                         & 1                                                                   & 1                                                                 & 1                                                                  & 1                                                                  & 1                                                                  & \textbf{9}                                                                \\ \hline
2   & Password in PBE                   & 3                                                               & 1                                                                       & 1                                                                         & 1                                                                   & 1                                                                 & 1                                                                  & 2                                                                  & 1                                                                  & \textbf{11}                                                                \\ \hline
3   & Password in KeyStore              & 2                                                               & 1                                                                       & 1                                                                         & 1                                                                   & 1                                                                 & 1                                                                  & 2                                                                  & 1                                                                  & \textbf{10}                                                                \\ \hline
4   & Hostname Verifier                 & 2                                                               & 0                                                                       & 0                                                                         & 0                                                                   & 0                                                                 & 0                                                                  & 0                                                                  & 0                                                                  & \textbf{2}                                                                 \\ \hline
5   & Certificate Validation            & 3                                                               & 0                                                                       & 0                                                                         & 0                                                                   & 0                                                                 & 0                                                                  & 0                                                                  & 0                                                                  & \textbf{3}                                                                 \\ \hline
6   & SSL Socket                        & 1                                                               & 0                                                                       & 0                                                                         & 0                                                                   & 0                                                                 & 0                                                                  & 0                                                                  & 0                                                                  & \textbf{1}                                                                 \\ \hline
7   & HTTP Protocol                     & 2                                                               & 1                                                                       & 1                                                                         & 1                                                                   & 1                                                                 & 1                                                                  & 0                                                                  & 1                                                                  & \textbf{8}                                                                 \\ \hline
8   & PRNG     & 2                                                               & 0                                                                       & 0                                                                         & 0                                                                   & 0                                                                 & 0                                                                  & 0                                                                  & 0                                                                  & \textbf{2}                                                                 \\ \hline
9   & Seed in PRNG                      & 3                                                               & 2                                                                       & 2                                                                         & 2                                                                   & 2                                                                 & 2                                                                  & 2                                                                  & 2                                                                  & \textbf{17}                                                                \\ \hline
10  & Salt in PBE & 2                                                               & 1                                                                       & 1                                                                         & 1                                                                   & 1                                                                 & 1                                                                  & 1                                                                  & 1                                                                  & \textbf{9}                                                                 \\ \hline
11  & Mode of Operation                 & 2                                                               & 1                                                                       & 1                                                                         & 1                                                                   & 1                                                                 & 1                                                                  & 0                                                                  & 1                                                                  & \textbf{8}                                                                 \\ \hline
12  & Initialization Vector             & 2                                                               & 1                                                                       & 1                                                                         & 1                                                                   & 1                                                                 & 1                                                                  & 2                                                                  & 1                                                                  & \textbf{10}                                                                \\ \hline
13  & Iteration in PBE            & 2                                                               & 1                                                                       & 1                                                                         & 1                                                                   & 1                                                                 & 1                                                                  & 1                                                                  & 1                                                                  & \textbf{9}                                                                 \\ \hline
14  & Symmetric Ciphers                 & 6                                                               & 5                                                                       & 5                                                                         & 5                                                                   & 5                                                                 & 5                                                                  & 0                                                                  & 5                                                                  & \textbf{36}                                                                \\ \hline
15  & Asymmetric Ciphers                & 1                                                               & 1                                                                       & 1                                                                         & 0                                                                   & 1                                                                 & 1                                                                  & 0                                                                  & 1                                                                  & \textbf{6}                                                                 \\ \hline
16  & Cryptographic Hash       & 5                                                               & 4                                                                       & 4                                                                         & 4                                                                   & 4                                                                 & 4                                                                  & 0                                                                  & 4                                                                  & \textbf{29}   \\ \hline
17  & Cryptographic MAC       & 3                                                               & 0                                                                       & 0                                                                         & 0                                                                   & 0                                                                 & 0                                                                  & 0                                                                  & 0                                                                  & \textbf{3}                                                              \\ \hline
18  & Credentials in String       & 2                                                              & 1                                                                       & 1                                                                         & 1                                                                   & 1                                                                 & 0                                                                  & 1                                                                  & 1                                                                  & \textbf{8}                                                              \\ \hline 
\multicolumn{2}{|c|}{\textbf{Total Cases per Group}}      & \textbf{45}                                                     & \textbf{21}                                                             & \textbf{21}                                                               & \textbf{20}                                                         & \textbf{21}                                                       & \textbf{20}                                                        & \textbf{12}                                                        & \textbf{21}                                                        & \textbf{181}                                                               \\ \hline
\end{tabular}
\vspace{-4mm}
\end{table*}

\vspace{1mm}
\textbf{\noindent{\em 2.13 Iteration Count in Password-based Encryption (PBE):}}
In \texttt{javax.crypto.spec.PBEParameterSpec} API, it takes iteration count as one of the parameters for Password-based Encryption (PBE). In PKCS \#5~\cite{moriarty2017pkcs}, it is suggested that the number of iteration should be more than 1000 to provide a reasonable security level.  

\vspace{1mm}
\textbf{\noindent{\em 2.14 Symmetric Ciphers:}}
In symmetric cryptography, the same key is used for encryption and decryption. Some symmetric ciphers, e.g., DES, Blowfish, RC4, RC2, IDEA are considered broken, as brute-force attack is possible for 64-bit ciphers. To overcome the attack, developers need to use AES which can support a block length of 128 bits and key lengths of 128, 192, and 256 bits~\cite{AESEncryption}. 



\vspace{1mm}
\textbf{\noindent{\em 2.15 Asymmetric Ciphers:}}
In asymmetric cryptography, two keys, i.e., a public key and a private key are used for encryption and decryption.  RSA is considered insecure for 1024-bit ciphers~\cite{244246}. For this reason, developers are recommended to use RSA with a key size of 2048 bits or higher. 

\vspace{1mm}
\textbf{\noindent{\em 2.16 Cryptographic Hash Functions:}}
 A cryptographic hash function generates a fixed-length alphanumeric hash value or message digest which is commonly used in verifying message integrity, digital signature, and authentication. A cryptographic hash function is contemplated as broken if a collision can be observed, i.e., the same hash value is generated for two different inputs. The list of broken hash functions includes SHA1, MD4, MD5, and MD2. Therefore, developers need to use a strong hash function, e.g., SHA-256. 


\vspace{1mm}
\textbf{\noindent{\em 2.17 Cryptographic MAC:}}
A  MAC algorithm HmacMD5 and HmacSHA1 are considered insecure as these are susceptible to collision attacks~\cite{bellare2015new}. Therefore, the developers need to use a strong MAC algorithm, e.g., HmacSHA256.

\vspace{1mm}
\textbf{\noindent{\em 2.18 Credentials in String:}}
Credentials (passwords, secret keys, etc) should not be stored in the String variable. In Java, String is a final and immutable class stored in the heap. More specifically, it exists in the memory until garbage collection. Therefore, sensitive information should not be stored in String\cite{stringimmutable, secureCode}. Compared with String, it is highly recommended to use mutable data structures (e.g., byte or char array) for sensitive information and clear it immediately after use. This reduces the window of opportunity for an adversary.~\cite{passwordmutable}.  

\section{Design of CryptoAPI-Bench}
\label{sec:design}
In this section, we present the design of the CryptoAPI-Bench. We manually generate 181 unit test cases guided by 18 types of misuses presented in Section~\ref{sec:threat_model}. We divide all test cases into two groups, i.e., basic cases and advanced cases. These test cases incorporate the majority of possible variations in the perspective of program analysis to detect cryptographic vulnerability. 
\label{sec:test_cases}

\vspace{-3mm}
\subsection{Basic Cases}
Basic test cases are simple ones where the probable source of vulnerability for Crypto API exists within the same method.  For example, Listing 1 shows that Cipher API takes \texttt{cryptoAlgo} as an argument. Note that, \texttt{cryptoAlgo} contains an insecure cipher algorithm that is defined within the same method \texttt{method1}.  In CryptoAPI-Bench, we create 45 basic test cases covering all 18 misuse categories. Among these test cases, 30 test cases contain cryptographic vulnerability (i.e., true positive), and 15 test cases do not contain any cryptographic vulnerability (i.e., true negative). These test cases identify a tool's capability to detect a specific misuse category.  

\begin{lstlisting}[language=Java, caption = Example code snippet of a basic test case, escapeinside=`', label = listing_basic]
 public void method1 () 
 {  ...
    cryptoAlgo = "DES/ECB/PKCS5Padding"
    Cipher cipher = Cipher.getInstance(`\underline{cryptoAlgo}')  
    ...
 }
\end{lstlisting}

\vspace{-3mm}
\subsection{Advanced Cases}
The advanced cases are more complex compared to basic cases where the probable source of vulnerability of a Crypto API appears from other methods, classes, field variables, or conditional statements . In CryptoAPI-Bench, we include 136 advanced cases. The distribution of advanced cases is presented from the fourth to tenth columns of TABLE~\ref{table:rule_vs_group}.   


\vspace{-3mm}
\subsubsection{Interprocedural Cases}

In interprocedural cases, the probable source of vulnerability in a Crypto API comes from other methods (i.e., procedures). We create two types of interprocedural cases: two-interprocedural (i.e., involving two methods) and three-interprocedural (i.e., involving three methods). In a two-interprocedural test case, the probable source of vulnerability comes from another method as a parameter.   Listing~\ref{listing_interprocedural} shows the code snippet of a two-interprocedural test case. In \texttt{method2},  Cipher API takes \texttt{cryptoAlgo} as an argument, and \texttt{cryptoAlgo} is not defined in \texttt{method2}, rather, it comes from another method \texttt{method1}. The assigned value of \texttt{cryptoAlgo} in \texttt{method1} shows that the test case is insecure. In three-interprocedural test cases, the probable source of vulnerability comes from two consecutive methods (i.e., source defined in one method, passes to another method, and then passes again to be used in Cipher API). CryptoAPI-Bench contains a total of 42 interprocedural test cases. Among them, 21 are two-interprocedural test cases, and 21 are three-interprocedural test cases.  The purpose of having the interprocedural test cases is to check the detection tool's interprocedural data flow handling capability.
\\

\begin{lstlisting}[language=Java, caption = Example code snippet of a two-interprocedural test case, escapeinside=`', label = listing_interprocedural]
 public void method1 () 
 {  ...
    cryptoAlgo = "DES/ECB/PKCS5Padding" 
    method2(cryptoAlgo)
    ...
 }
 public void method2 (String `\underline{cryptoAlgo}') 
 {  ...
    Cipher cipher = Cipher.getInstance(`\underline{cryptoAlgo}')  
    ...
 }
\end{lstlisting}

\vspace{-3mm}
\subsubsection{Field Sensitive Cases}
In field-sensitive cases, the probable source of cryptographic vulnerabilities can be detected by the analysis tools if the tools are capable of performing field-sensitive data flow analysis. Field-sensitive refers to an analysis that is able to differentiate multiple fields or variables with the same object~\cite{10.1145/3290361}. In Listing~\ref{listing_field_sensitive}, \texttt{algo} is an instance or field variable in the \texttt{Crypto} class. The constructor \texttt{Crypto()} stores \texttt{algo} with defAlgo object. A class member function \texttt{encrypt()} use this \texttt{algo} value in Cipher API. Both \texttt{algo} and \texttt{defAlgo} contain the same object, i.e., a secure or insecure cipher algorithm. This is a field-sensitive case as the tools need to trace the field variable \texttt{algo} as the probable source of vulnerability. CryptoAPI-Bench contains 20 field-sensitive test cases.

\begin{lstlisting}[language=Java, caption = Example code snippet of a field sensitive test case, escapeinside=`', label = listing_field_sensitive]
class Crypto {
    String `\underline{algo}'
    public Crypto (String `\underline{defAlgo}') {
        algo = `\underline{defAlgo}';
    }
    public void encrypt(... ) {
        ...
        Cipher cipher = Cipher.getInstance(`\underline{algo}');
        ...
    }
 }
\end{lstlisting}

\vspace{-3mm}
\subsubsection{Combined Cases}
The combined cases are a bit more complex where both interprocedural and field sensitivity properties are combined, i.e., both Listing~\ref{listing_interprocedural} and Listing~\ref{listing_field_sensitive} are incorporated to generate complicated test cases. CryptoAPI-Bench has 21 combined test cases.
\vspace{-3mm}
\subsubsection{Path-Sensitive Cases}
In path-sensitive test cases, conditional branch instructions are included in the test cases containing the definition of the probable source of a vulnerability. In Listing~\ref{listing_path_sensitive}, an example code snippet of a path sensitivity case is given. Depending on the \texttt{choice} variable, the \texttt{Cipher} is getting the instance from a secure or an insecure cryptographic algorithm. There are 20 path-sensitive test cases in CryptoAPI-Bench.  


\begin{lstlisting}[language=Java, caption = Example code snippet of a path sensitive test case, escapeinside=`', label = listing_path_sensitive]
public void method1 (int `\underline{choice}') {
    ...
    Cipher ch = Cipher.getInstance (`\underline{"DES/ECB/..."}') ;
    if (`\underline{choice}' > 1) {
        ch = Cipher.getInstance (`\underline{"AES/CBC/..."}') ;
    }
    `\underline{ch}'.init (Cipher.ENCRYPT_MODE, key) ;
    ...
 }
\end{lstlisting}
\vspace{-3mm}
\subsubsection{Miscellaneous Cases}
Miscellaneous test cases evaluate the tool's abilities to recognize irrelevant constraints and other interfaces, e.g., Map.  In Listing~\ref{listing_false_positive}, the Map interface of Line 3-6 provides a secure key or insecure key depending on the choice variable. The Map indices (e.g., ``a", ``b") represent only index values, not security-relevant values. Similarly, in Line 8, the ``UTF-8" represents byte encoding, not any constant or hard-coded value. CryptoAPI-Bench contains 12 miscellaneous test cases. 


\begin{lstlisting}[language=Java, caption = Example code snippet of a miscellaneous test case, escapeinside=`', label = listing_false_positive]
public void method1 (String `\underline{choice}') {
    ...
    Map<String,String> hm = new HashMap<String, String>();
    hm.put("a", secureKeyString);
    hm.put("b", insecureKeyString);
    String keyString = hm.get(`\underline{choice}');
    
    byte [] b = secureKeyString.getBytes(`\underline{"UTF-8"}');
    IvParameterSpec ivSpec = new IvParameterSpec(b);
    ...
}
\end{lstlisting}

\vspace{-3mm}
\subsubsection{Multiple Class Cases}
In multiple class test cases, the probable source of vulnerabilities comes from another Java class.  An example code snippet of multiple class case is presented in Listing~\ref{listing_multiple_class}. It is necessary to detect whether a secure or an insecure algorithm is passed in Line 4 in \texttt{MultipleClass1} and used in Line 9 in \texttt{MultipleClass2}. CryptoAPI-Bench has 21 multiple class test cases.

\begin{lstlisting}[language=Java, caption = Example code snippet of a multiple class test case, escapeinside=`', label = listing_multiple_class]
public class MultipleClass1 { 
    public void method1 (String `\underline{passedAlgo}') {
        MultipleClass2 mc = new MultipleClass2 ();
        mc.method2 (passedAlgo);
    }
}
public class MultipleClass2 { 
    public void method2 (String `\underline{cryptoAlgo}') {
        Cipher c = Cipher.getInstance (`\underline{cryptoAlgo}');  
    }
}
\end{lstlisting}

\vspace{-4mm}
\section{Design of \apachebenchname{}}
\label{sec:design_apache}
We include the early version of real-world large 10 Apache projects to check the scalability property of different tools. The second and third columns of TABLE~\ref{tab:apachebench_stat} show the number of Java files and lines of Java Code in Apache projects. The spark project is the largest among 10 considered projects containing 2,005 Java files with 311,856 lines of code.  The meecrowave project contains the lowest number of Java files (40 Java files) and deltaspike contains the lowest number of lines of code (i.e., 5,116 LoC).

\begin{table*}[]
\centering
\caption{\apachebenchname{}: Summary of unit test cases. Contents (number of Java file and lines of code) of the considered Apache projects are summarized here. There are total 121 unit test cases with 79 basic cases and 42 advanced cases. Details information are presented in Section~\ref{sec:design_apache}. }
\vspace{-3mm}
\label{tab:apachebench_stat}
\begin{tabular}{|l|c|c|c|c|c|c|c|}
\hline
\rowcolor[HTML]{C0C0C0} 
\cellcolor[HTML]{C0C0C0}                                                                                     & \cellcolor[HTML]{C0C0C0}                                                                                           & \cellcolor[HTML]{C0C0C0}                                                                                    & \multicolumn{5}{c|}{\cellcolor[HTML]{C0C0C0}\textbf{Test Cases}}                                                      \\ \cline{4-8} 
\rowcolor[HTML]{C0C0C0} 
\multirow{-2}{*}{\cellcolor[HTML]{C0C0C0}\textbf{\begin{tabular}[c]{@{}l@{}}Apache \\ Project\end{tabular}}} & \multirow{-2}{*}{\cellcolor[HTML]{C0C0C0}\textbf{\begin{tabular}[c]{@{}c@{}}Number of \\ Java Files\end{tabular}}} & \multirow{-2}{*}{\cellcolor[HTML]{C0C0C0}\textbf{\begin{tabular}[c]{@{}c@{}}Lines of \\ Code\end{tabular}}} & \textbf{Total Case} & \textbf{Basic Case} & \textbf{Advanced Cases} & \textbf{True Positive} & \textbf{True Negative} \\ \hline
deltaspike                                                                                                   & 87                                                                                                                 & 5116                                                                                                        & 5                   & 2                   & 3                       & 2                      & 3                      \\ \hline
directory-server                                                                                             & 468                                                                                                                & 20780                                                                                                       & 36                  & 15                  & 21                      & 19                     & 17                     \\ \hline
incubator-taverna-workbench                                                                                  & 45                                                                                                                 & 9919                                                                                                        & 8                   & 5                   & 3                       & 5                      & 3                      \\ \hline
manifoldcf                                                                                                   & 126                                                                                                                & 16998                                                                                                       & 7                   & 4                   & 3                       & 5                      & 2                      \\ \hline
meecrowave                                                                                                   & 40                                                                                                                 & 5646                                                                                                        & 3                   & 3                   & 0                       & 3                      & 0                      \\ \hline
spark                                                                                                        & 2005                                                                                                               & 311856                                                                                                      & 27                  & 27                  & 0                       & 27                     & 0                      \\ \hline
tika                                                                                                         & 225                                                                                                                & 16558                                                                                                       & 3                   & 0                   & 3                       & 0                      & 3                      \\ \hline
tomee                                                                                                        & 1029                                                                                                               & 118661                                                                                                      & 9                   & 6                   & 3                       & 7                      & 2                      \\ \hline
wicket                                                                                                       & 204                                                                                                                & 13442                                                                                                       & 8                   & 4                   & 4                       & 5                      & 3                      \\ \hline
artemis-commons                                                                                              & 126                                                                                                                & 8915                                                                                                        & 15                  & 13                  & 2                       & 15                     & 0                      \\ \hline
\multicolumn{3}{|r|}{\textbf{Total}}                                                                                                                                                                                                                                                                                                            & \textbf{121}        & \textbf{79}         & \textbf{42}             & \textbf{88}            & \textbf{33}            \\ \hline
\end{tabular}
\vspace{-4mm}
\end{table*}

We enlist 121 test cases in \apachebenchname{}. Among them, 79 test cases are basic cases, i.e., the vulnerability rise within the same method. There are 42 advanced test cases where probable source vulnerability comes from other methods (interprocedural cases), other classes (multiple class cases), class variables (field sensitive cases), etc. We detect 88 cryptographic misuses, i.e., true positive alerts. Regarding true negatives, we consider only the cases where a tool shows the case as a false alert. With this consideration, we show 33 true negative cases.  

We look into the Apache projects in the Benchmark and made detailed documentation. The documentation consists of cryptographic vulnerabilities the project contains, an explanation of the error, the location (file name, method name, line number) of the vulnerabilities. The documentation and corresponding \apachebenchname{} benchmark are available in the GitHub repository~\cite{apachecryptoapibench-github}.

\vspace{-3mm}
\section{Existing Cryptographic Vulnerability Detection Tools}
\label{sec:existing_tools}

In this section, we summarize the vulnerability detection tools that we choose to run on CryptoAPI-Bench and \apachebenchname{}. We consider three criteria while choosing the analysis tools. 
(1) Open-sourced tools: The open-sourced vulnerability detection tools, i.e., CrySL~\cite{DBLP:conf/ecoop/KrugerS0BM18}, CryptoGuard~\cite{DBLP:journals/corr/abs-1806-06881}, SpotBugs~\cite{spotbugs} are convenient to use as we are able to analyze their codes and understand the reason for their lack of performance. 
(2) Static analysis tools: We choose static analysis tools that can examine and detect vulnerability without executing the code. SpotBugs, CryptoGuard, CrySL, and Coverity~\cite{coverity} are static analysis tools. 
(3) Free cryptographic vulnerability detection services:  We consider Coverity as a provider of free cryptographic vulnerability detection service. Coverity is not open-sourced. However, Coverity provides online services to detect vulnerability.

\begin{table*}[]
\centering
\scriptsize
\caption{Generated alert keywords for each misuse category from cryptographic vulnerability detection tools (SpotBugs, CryptoGuard, CrySL, and Coverity). For example, for misuse category 16 (i.e., Cryptographic Hash), the generated alert keywords in tools are WEAK\_MESSAGE\_DIGEST, broken hash scheme, ConstraintError, RISKY\_CRYPTO, respectively.}
\vspace{-3mm}
\label{table:crypto_alert_keywords}
\begin{tabular}{|c|c|c|c|c|}
\hline
\rowcolor{lightgray} 
\textbf{Misuse Categories} & \textbf{SpotBugs}        & \textbf{CryptoGuard}            & \textbf{CrySL}         & \textbf{Coverity}         \\ \hline 
  1     & HARD\_CODE\_PASSWORD    & Constant keys                   & RequiredPredicateError & HARDCODED\_CREDENTIALS    \\ 
2     & HARD\_CODE\_PASSWORD    & Constant keys                   & HardCodedError       & HARDCODED\_CREDENTIALS    \\ 
3     & HARD\_CODE\_PASSWORD    & Predictable password            & HardCodedError       & HARDCODED\_CREDENTIALS    \\ 
4     & WEAK\_HOSTNAME\_VERIFIER & Manually verify hostname        & ---                    & BAD\_CERT\_VERIFICATION   \\ 
5     & WEAK\_TRUST\_MANAGER     & Untrusted TrustManager          & ---                     & BAD\_CERT\_VERIFICATION   \\ 
6     & --                       & Does not manually verify socket & ---        & RESOURCE\_LEAK            \\ 
7     & ---                      & HTTP protocol                   & ---                    & ---                       \\ 
8     & PREDICTABLE\_RANDOM      & Untrusted PRNG                  & RequiredPredicateError                     & ---                       \\ 
9     & ---                      & Predictable Seed                & RequiredPredicateError          & PREDICTABLE\_RANDOM\_SEED \\ 
10    & ---                      & Constant Salt                   & RequiredPredicateError & ---                       \\ 
11    & CIPHER\_INTEGRITY        & Broken crypto scheme            & ConstraintError        & RISKY\_CRYPTO             \\ 
12    & STATIC\_IV               & Constant IV                     & RequiredPredicateError & ---                       \\ 
13    & ---                      & \textless 1000 iteration        & ConstraintError        & ---                       \\ 
14    & CIPHER\_INTEGRITY        & Broken crypto scheme            & ConstraintError        & RISKY\_CRYPTO             \\ 
15    & ---                      & Export grade public key         & ConstraintError        & ---                       \\ 
16    & WEAK\_MESSAGE\_DIGEST    & Broken hash scheme              & ConstraintError        & RISKY\_CRYPTO             \\ 
17    & ---    & ---              & ConstraintError        & ---  \\ 
18    & ---    & ---              & RequiredPredicateError        & ---             \\\hline 
\end{tabular}
\vspace{-4mm}
\end{table*}

We also consider GrammaTech~\cite{grammatech}, QARK~\cite{qark} and FixDroid~\cite{DBLP:conf/ccs/NguyenWABWF17}. However, GrammaTech is a commercial tool. We were unable to access its trial version. The online SWAMP~\cite{swamp} contains GrammaTech tool to use that only supports vulnerability detection for C and C++. Therefore, we excluded GrammaTech from our list of tools. QARK is a tool that is mainly designed to capture security vulnerabilities in Android applications. FixDroid is built as a research prototype that is embedded as a plugin in Android Studio to conduct a usability study. Our investigation shows that the detection capability of FixDroid and QARK is limited. Though QARK has been maintained and updated, FixDroid has not been updated since 2017.

Therefore, we mainly focus on four tools, i.e., SpotBugs, CryptoGuard, CrySL, and Coverity to evaluate on CryptoAPI-Bench.

\vspace{-3mm}
\subsection{SpotBugs}
SpotBugs is a static analysis tool also for capturing deficiencies in Java code. The tool is built based on a plugin structure. The tools detect defects by utilizing visitor patterns in class files or bytecodes of Java, state machine, flags. We use the SpotBugs tool (version 3.1.12) available online in SWAMP~\cite{swamp}. However, currently, SWAMP is in the transition to a new host service~\cite{transitionswamp}.  

\vspace{-3mm}
\subsection{CryptoGuard}
CryptoGuard~\cite{DBLP:journals/corr/abs-1806-06881} is a static analysis tool that is operated based on program slicing with novel language-based refinement algorithms. It significantly reduces the false positive rate which is a typical problem for static analysis. Furthermore, CryptoGuard covers 16 cryptographic rules and achieves high precision. The authors showed screening a large number of Apache projects and Android apps to present their high precision rate and low false positive rate. We run the experiment on CryptoGuard (commitID:  97b220) available on GitHub~\cite{cryptoguardgithub}.

\vspace{-3mm}
\subsection{CrySL}
CrySL~\cite{DBLP:conf/ecoop/KrugerS0BM18} is a domain-specific language for cryptographic libraries. The static analysis $CogniCrypt_{SAST}$ takes the rules provided in the specification language CrySL as input, and performs a static analysis based on the specification of the rules. CrySL is open-sourced and we run the experiment on CrySL (commit  ID:  004cd2) available on GitHub~\cite{cryslgithub}.


\vspace{-3mm}
\subsection{Coverity}
Coverity is a commercial tool that analyzes the vulnerabilities of codes. Unlike other tools, it takes the source code and performs its analysis. The Coverity analysis tool is available to use online~\cite{coverity}. We perform the latest analysis using Coverity around September 2020.

\vspace{-4mm}
\section{Evaluation and Analysis}
\label{sec:evaluation_analysis}

\begin{table*}[h]
\caption{CryptoAPI-Bench comparison of SpotBugs, CryptoGuard, CrySL and Coverity on all 18 rules with CryptoAPI-Bench's 181 test cases. There are 37 secure API use cases (15 in basic and 22 in advanced), which a tool should not raise any alerts on. GTP stands for ground truth positive, which is the number of insecure API use cases in the benchmark. Findings of the table are reported in Section~\ref{sec:analysis_overview}.}
\vspace{-3mm}
\label{table:totalbench}
\centering
\begin{tabular}{|c|l|l|cc|cc|cc|cc|}
\hline \rowcolor{lightgray} 
{No.} & \multicolumn{1}{c|}{\textbf{Misuse Categories}} & \multicolumn{1}{c|}{\textbf{GTP}} & \multicolumn{2}{c|}{\textbf{SpotBugs}} & \multicolumn{2}{c|}{\textbf{CryptoGuard}} & \multicolumn{2}{c|}{\textbf{CrySL}} & \multicolumn{2}{c|}{\textbf{Coverity}} \\ \cline{4-11} \rowcolor{lightgray} 
                         & \multicolumn{1}{c|}{} & \multicolumn{1}{c|}{}                       & \textbf{TP}           & \textbf{FP}          & \textbf{TP}             & \textbf{FP}           & \textbf{TP}            & \textbf{FP}            & \textbf{TP}              & \textbf{FP}             \\ \hline  
1                        & Cryptographic Key & 7              & 0            & 3           & 5              & 1            & 0             & 8             & 5               & 1              \\ 
2                        & Password in PBE & 8               & 2            & 0           & 7              & 1            & 0             & 10             & 7               & 1              \\ 
3                        & Password in KeyStore & 7          & 1            & 1           & 7              & 1            & 0             & 10             & 5               & 1              \\ 
4                        & Hostname Verifier & 1                    & --           & --          & 1              & 0            & --             & --             & 1               & 0              \\ 
5                        & Certificate Validation  & 3                    & 3           & 0          & 3              & 0            & --             & --             & 3               & 0              \\ 
6                        & SSL Socket & 1                       & --           & --          & 1              & 0            & --            & --            & 1               & 0              \\ 
7                        & HTTP Protocol  & 6                               & --           & --          & 6             & 1           & --            & --            &  --               & --              \\ 
8                        & PRNG & 1                            & 1           & 0          & 1              & 0            & 1            & 0            & --               & --              \\ 
9                        & Seed in PRNG & 14                             & --           & --          & 11             & 2           & 0             & 15             & 1               & 2              \\ 
10                       & Salt in PBE  & 7                               & --            & --           & 6             & 1           & 6            & 1            & --               & --              \\ 
11                       & Mode of Operation & 6                         & 1            & 3           & 6              & 1            & 5             & 1             & 1               & 1              \\ 
\multicolumn{1}{|l|}{12} & Initialization Vector  & 8                                   & 3            & 6           & 7             & 1           & 7             & 1             & --               & --              \\ 
\multicolumn{1}{|l|}{13} & Iteration Count in PBE &  7               & --            & --           & 5             & 1           & 5            & 3            & --               & --              \\ 
\multicolumn{1}{|l|}{14} & Symmetric Cipher  & 30                       & 5           & 11           & 30              & 5            & 25             & 5             & 4              & 4              \\ 
\multicolumn{1}{|l|}{15} & Asymmetric Ciphers  & 5                    & --            & --           & 4             & 1           & 5             & 1             & --               & --              \\ 
\multicolumn{1}{|l|}{16} & Cryptographic Hash  & 24                               & 4            & 8           & 24              & 4            & 20             & 4             & 4              & 4              \\
\multicolumn{1}{|l|}{17} & MAC Algorithm  & 2                               & --            & --           & --              & --            & 2             & 0             & --              & --              \\
\multicolumn{1}{|l|}{18} & Credentials in String   & 7                               & --            & --           & --              & --            & 0             & 8             & --              & --              \\  \hline 
\multicolumn{2}{|c|}{\textbf{Total}} & \textbf{144}                                   & \textbf{20}  & \textbf{32} & \textbf{124}    & \textbf{20}   & \textbf{76}   & \textbf{67}   & \textbf{32}     & \textbf{14}     \\ \hline
\end{tabular}
\vspace{-4mm}
\end{table*}

In this section, we evaluate the results for four cryptographic misuse detection tools, i.e., SpotBugs, CryptoGuard, CrySL and Coverity. We show the experimental setup, evaluation criteria, and analysis results using both benchmarks.


\vspace{-3mm}
\subsection{Experimental Setup}
We evaluate mainly four cryptographic analysis tools, i.e., SpotBugs~\cite{spotbugs}, CryptoGuard~\cite{DBLP:journals/corr/abs-1806-06881}, CrySL~\cite{DBLP:conf/ecoop/KrugerS0BM18}, Coverity~\cite{coverity} on both Benchmarks. We follow the instructions from GitHub to set up the environment of CryptoGuard and CrySL in our machine to perform the analysis. We upload JAR files from CryptoAPI-Bench and Apache projects into SpotBugs tool available in SWAMP.  Coverity is an online commercial tool~\cite{coverity} that takes GitHub link and compressed code files in order to start analysis execution. 



\vspace{-3.5mm}
\subsection{Evaluation Criteria}
\label{sec:evaluation_criteria}
We evaluate the vulnerability detection tools by running these tools on our benchmarks. After performing the analysis, we capture true positives, false positives, and false negatives from the corresponding tool's result log. As our purpose is to detect cryptographic vulnerability detection, we consider only cryptographic misuse alerts and discard others. In TABLE~\ref{table:crypto_alert_keywords}, we present the alert keywords that detection tools use while showing a specific cryptographic misuse. This can assist developers to understand which keyword they should search in the result log to find a specific type of vulnerability. In the following, we provide a brief description of our process of identification of true positive, false positive, and false negative alerts.     
\vspace{-2mm}
\subsubsection{True positive (TP)}
If a tool generates an alert due to the correct reason while screening any specific vulnerable unit test case in CryptoAPI-Bench, then the event is considered as a true positive. 

\vspace{-2mm}

\vspace{-0.5mm}
\subsubsection{False positive (FP)}
The false positive alert can be captured from two different scenarios. If an alert raised by a tool is unexpected (i.e., does not exist in a specific unit test case), then the alert is a false positive. In addition, if a tool gives an inaccurate reason for an expected alert, then it is also considered a false positive. 

\vspace{-2mm}

\subsubsection{False negative (FN)}
A vulnerable test case may not be detected by the evaluation tools. This missed detection is characterized as a false negative.

After analyzing the results by determining the true positive (TP), false positive (FP), and false negative (FN) values, we compute the recall and precision to determine the performance of the tools. 

\begin{table*}[]
\centering
\label{table:basic_table}
\caption{CryptoAPI-Bench comparison of SpotBugs, CryptoGuard, CrySL and Coverity on six common misuse categories with CryptoAPI-Bench's common 21 basic cases. TP, FP, FN stand for true positive, false positive, false negative, respectively. Findings of the table are reported in Section~\ref{sec:basic_findings}.}
\vspace{-3mm}
\begin{tabular}{|l|c|c|c|l|l|c|l|l|c|l|l|c|l|l|}
\hline
\rowcolor[HTML]{C0C0C0} 
\cellcolor[HTML]{C0C0C0}                                            & \cellcolor[HTML]{C0C0C0}                                                                                          & \cellcolor[HTML]{C0C0C0}                                                                                          & \multicolumn{3}{c|}{\cellcolor[HTML]{C0C0C0}\textbf{SpotBugs}}                                                                                & \multicolumn{3}{c|}{\cellcolor[HTML]{C0C0C0}\textbf{CryptoGuard}}                                                                             & \multicolumn{3}{c|}{\cellcolor[HTML]{C0C0C0}\textbf{CrySL}}                                                                                   & \multicolumn{3}{c|}{\cellcolor[HTML]{C0C0C0}\textbf{Coverity}}                                                                                \\ \cline{4-15} 
\rowcolor[HTML]{C0C0C0} 
\multirow{-2}{*}{\cellcolor[HTML]{C0C0C0}\textbf{Basic Test Cases}} & \multirow{-2}{*}{\cellcolor[HTML]{C0C0C0}\textbf{\begin{tabular}[c]{@{}c@{}}True Positive \\ Count\end{tabular}}} & \multirow{-2}{*}{\cellcolor[HTML]{C0C0C0}\textbf{\begin{tabular}[c]{@{}c@{}}True Negative \\ Count\end{tabular}}} & \textbf{TP}             & \multicolumn{1}{c|}{\cellcolor[HTML]{C0C0C0}\textbf{FP}} & \multicolumn{1}{c|}{\cellcolor[HTML]{C0C0C0}\textbf{FN}} & \textbf{TP}             & \multicolumn{1}{c|}{\cellcolor[HTML]{C0C0C0}\textbf{FP}} & \multicolumn{1}{c|}{\cellcolor[HTML]{C0C0C0}\textbf{FN}} & \textbf{TP}             & \multicolumn{1}{c|}{\cellcolor[HTML]{C0C0C0}\textbf{FP}} & \multicolumn{1}{c|}{\cellcolor[HTML]{C0C0C0}\textbf{FN}} & \textbf{TP}             & \multicolumn{1}{c|}{\cellcolor[HTML]{C0C0C0}\textbf{FP}} & \multicolumn{1}{c|}{\cellcolor[HTML]{C0C0C0}\textbf{FN}} \\ \hline
IntraProcedural                                                     & 14                                                                                                                & 6                                                                                                                 & \multicolumn{1}{l|}{13} & 3                                                        & 1                                                        & \multicolumn{1}{l|}{14} & 0                                                        & 0                                                        & \multicolumn{1}{l|}{10} & 7                                                        & 4                                                        & \multicolumn{1}{l|}{13} & 0                                                        & 1                                                        \\ \hline
\multicolumn{2}{|c|}{}                                                                                                                                                                  & \textbf{Recall (\%)}                                                                                              & \multicolumn{3}{c|}{\textbf{92.86}}                                                                                                           & \multicolumn{3}{c|}{\textbf{100.00}}                                                                                                          & \multicolumn{3}{c|}{\textbf{71.43}}                                                                                                           & \multicolumn{3}{c|}{\textbf{92.86}}                                                                                                           \\ \cline{3-15} 
\multicolumn{2}{|c|}{\multirow{-2}{*}{\textbf{Result}}}                                                                                                                                 & \textbf{Precision (\%)}                                                                                           & \multicolumn{3}{c|}{\textbf{81.25}}                                                                                                           & \multicolumn{3}{c|}{\textbf{100.00}}                                                                                                          & \multicolumn{3}{c|}{\textbf{58.82}}                                                                                                           & \multicolumn{3}{c|}{\textbf{100.00}}                                                                                                          \\ \hline
\end{tabular}
\vspace{-4mm}
\end{table*}

\begin{table*}[]
\centering
\caption{CryptoAPI-Bench comparison of SpotBugs, CryptoGuard, CrySL and Coverity on six common misuse categories with CryptoAPI-Bench's common 84 advanced cases. TP, FP, FN stand for true positive, false positive, false negative, respectively. Findings of the table are reported in Section~\ref{sec:advanced_findings}.}
\vspace{-3mm}
\label{table:advanced}
\begin{tabular}{|l|c|c|c|c|c|c|c|c|c|c|c|c|c|c|}
\hline
\rowcolor[HTML]{C0C0C0} 
\multicolumn{1}{|c|}{\cellcolor[HTML]{C0C0C0}}                                                                                          & \cellcolor[HTML]{C0C0C0}                                                                                          & \cellcolor[HTML]{C0C0C0}                                                                                          & \multicolumn{3}{c|}{\cellcolor[HTML]{C0C0C0}\textbf{SpotBugs}} & \multicolumn{3}{c|}{\cellcolor[HTML]{C0C0C0}\textbf{CryptoGuard}} & \multicolumn{3}{c|}{\cellcolor[HTML]{C0C0C0}\textbf{CrySL}} & \multicolumn{3}{c|}{\cellcolor[HTML]{C0C0C0}\textbf{Coverity}} \\ \cline{4-15} 
\rowcolor[HTML]{C0C0C0} 
\multicolumn{1}{|c|}{\multirow{-2}{*}{\cellcolor[HTML]{C0C0C0}\textbf{\begin{tabular}[c]{@{}c@{}}Advanced \\ Test Cases\end{tabular}}}} & \multirow{-2}{*}{\cellcolor[HTML]{C0C0C0}\textbf{\begin{tabular}[c]{@{}c@{}}True Positive \\ Count\end{tabular}}} & \multirow{-2}{*}{\cellcolor[HTML]{C0C0C0}\textbf{\begin{tabular}[c]{@{}c@{}}True Negative \\ Count\end{tabular}}} & \textbf{TP}         & \textbf{FP}         & \textbf{FN}        & \textbf{TP}          & \textbf{FP}          & \textbf{FN}         & \textbf{TP}        & \textbf{FP}        & \textbf{FN}       & \textbf{TP}         & \textbf{FP}         & \textbf{FN}        \\ \hline
Two-Interprocedural                                                                                                                     & 13                                                                                                                & 0                                                                                                                 & 0                   & 0                   & 13                 & 12                   & 0                    & 1                   & 10                 & 3                  & 3                 & 3                   & 0                   & 10                 \\ \hline
Three-Interprocedural                                                                                                                   & 13                                                                                                                & 0                                                                                                                 & 0                   & 0                   & 13                 & 12                   & 0                    & 1                   & 10                 & 3                  & 3                 & 3                   & 0                   & 10                 \\ \hline
Field Sensitive                                                                                                                         & 13                                                                                                                & 0                                                                                                                 & 0                   & 0                   & 13                 & 13                   & 0                    & 0                   & 10                 & 2                  & 3                 & 1                   & 0                   & 12                 \\ \hline
Combined Case                                                                                                                           & 13                                                                                                                & 0                                                                                                                 & 0                   & 12                  & 13                 & 12                   & 0                    & 1                   & 0                  & 2                  & 13                & 3                   & 0                   & 10                 \\ \hline
Path Sensitive                                                                                                                          & 0                                                                                                                 & 13                                                                                                                & 0                   & 10                  & 0                  & 0                    & 13                   & 0                   & 0                  & 13                 & 0                 & 0                   & 12                  & 0                  \\ \hline
Miscellaneous Cases                                                                                                                     & 3                                                                                                                 & 2                                                                                                                 & 0                   & 0                   & 3                  & 3                    & 0                    & 0                   & 0                  & 5                  & 3                 & 0                   & 0                   & 3                  \\ \hline
Multiple Class methods                                                                                                                  & 13                                                                                                                & 0                                                                                                                 & 0                   & 0                   & 13                 & 13                   & 0                    & 0                   & 10                 & 3                  & 3                 & 3                   & 0                   & 10                 \\ \hline
\multicolumn{2}{|c|}{}                                                                                                                                                                                                                                      & \textbf{Recall (\%)}                                                                                              & \multicolumn{3}{c|}{\textbf{0.00}}                             & \multicolumn{3}{c|}{\textbf{95.59}}                               & \multicolumn{3}{c|}{\textbf{58.82}}                         & \multicolumn{3}{c|}{\textbf{19.12}}                            \\ \cline{3-15} 
\multicolumn{2}{|c|}{\multirow{-2}{*}{\textbf{Results}}}                                                                                                                                                                                                    & \textbf{Precision (\%)}                                                                                           & \multicolumn{3}{c|}{\textbf{0.00}}                             & \multicolumn{3}{c|}{\textbf{83.33}}                               & \multicolumn{3}{c|}{\textbf{56.34}}                         & \multicolumn{3}{c|}{\textbf{52.00}}                            \\ \hline
\end{tabular}
\vspace{-4mm}
\end{table*}

\vspace{-3mm}
\subsection{CryptoAPI-Bench: Analysis of Results}
\label{sec:analysis_overview}
In this section, we describe CryptoAPI-Bench evaluation findings on each detection tool based on the result log and performance analysis. TABLE~\ref{table:totalbench} presents the number of true positive and false positive vulnerability threat detection captured by the tools for 18 cryptographic misuse categories. There are only 6 common cryptographic misuse categories detected by all tools. To ensure fairness in comparison, we consider only these 6 common cryptographic misuses while finding the comparative analysis results of tools based on the basic and advanced benchmark in TABLE~5 and TABLE~\ref{table:advanced}, respectively. The analysis results are presented in terms of false positive rate (FPR), false negative rate (FNR), recall, and precision.

\noindent
\textbf{Analysis Overview: }
TABLE~\ref{table:totalbench} shows that among the 18 specified high impact cryptographic misuse categories in Section~\ref{sec:threat_model}, the cryptographic vulnerability detection tools are able to detect a subset of rules. 
\begin{itemize}
    \item SpotBugs, CryptoGuard, CrySL, Coverity covers 9, 16, 14, 10  cryptographic misuse categories, respectively.
    \item In total, the benchmark contains 144 vulnerable test cases and among these true positive cases, SpotBugs, CryptoGuard, CrySL, Coverity detects 20, 124, 76, 32 cases, respectively.
    \item In addition, SpotBugs, CryptoGuard, CrySL, Coverity also generate 32, 20, 67, 14 false alarms, respectively that are included as false positive cases. 
\end{itemize}

\vspace{-3mm}
\subsubsection{Analysis on Basic Benchmark}
\label{sec:basic_findings}
TABLE~5 shows the performance analysis result of four detection tools on six common cryptographic misuse categories based on the basic benchmark. We capture the following findings based on TABLE~5.

\begin{itemize}
    \item SpotBugs does not produce any false positive errors. It detects all cases except one. SpotBugs is not designed to capture threats in the basic case of the vulnerable cryptographic key misuse.
    
    
    \item CrySL produces 7 false positive errors due to maintaining strict rules in Crypto APIs of the cryptographic key, password in PBE, password in KeyStore.  
    
    \item Coverity does not generate any false positive errors. It can successfully detect every vulnerability except one. Coverity is not designed to capture IDEA as a vulnerable cryptographic algorithm. 
    \item For insecure uses of pseudo-random number generators, SpotBugs and CryptoGuard flag all uses of java.util.Random. However, CrySL flags the insecure random variable when they use in crypto contexts.
\end{itemize}

In summary, for all basic cases, CryptoGuard and Coverity generate a precision of 100\%. For SpotBugs and CrySL, it produces some false positives and hence generates precision of 81.25\%, 58.82\% respectively. 

\vspace{-2mm}
\subsubsection{Analysis on Advanced Benchmark}
\label{sec:advanced_findings}
TABLE~\ref{table:advanced} shows the performance analysis result of four detection tools on six common cryptographic misuse categories based on the advanced benchmark. We capture the following findings based on TABLE~\ref{table:advanced}.

\begin{table*}[]
\centering
\caption{\apachebenchname{} comparison of SpotBugs, CryptoGuard, CrySL and Coverity on 10 Apache projects. GTP stands for ground truth positive, which is the number of insecure API use cases in the Apache codes.}
\vspace{-3mm}
\label{tab:apache_result}
\begin{tabular}{|l|c|c|c|c|c|c|c|c|c|c|c|c|c|}
\hline
\rowcolor[HTML]{C0C0C0} 
\cellcolor[HTML]{C0C0C0}                                          & \cellcolor[HTML]{C0C0C0}                               & \multicolumn{3}{c|}{\cellcolor[HTML]{C0C0C0}\textbf{SpotBugs}} & \multicolumn{3}{c|}{\cellcolor[HTML]{C0C0C0}\textbf{CryptoGuard}} & \multicolumn{3}{c|}{\cellcolor[HTML]{C0C0C0}\textbf{CrySL}} & \multicolumn{3}{c|}{\cellcolor[HTML]{C0C0C0}\textbf{Coverity}} \\ \cline{3-14} 
\rowcolor[HTML]{C0C0C0} 
\multirow{-2}{*}{\cellcolor[HTML]{C0C0C0}\textbf{Apache Project}} & \multirow{-2}{*}{\cellcolor[HTML]{C0C0C0}\textbf{GTP}} & \textbf{TP}         & \textbf{FP}         & \textbf{FN}        & \textbf{TP}          & \textbf{FP}          & \textbf{FN}         & \textbf{TP}        & \textbf{FP}        & \textbf{FN}       & \textbf{TP}         & \textbf{FP}         & \textbf{FN}        \\ \hline
deltaspike                                                                          & 2                                                                                & 2           & 0           & 0           & 2            & 0            & 0           & 2           & 3           & 0           & 2           & \multicolumn{1}{c|}{0}          & \multicolumn{1}{c|}{0}           \\ \hline
directory-server                                                                    & 19                                                                               & 11          & 0           & 8           & 5            & 0            & 14          & 18          & 17          & 1           & 5           & \multicolumn{1}{c|}{0}          & \multicolumn{1}{c|}{14}          \\ \hline
incubator-taverna-workbench                                                         & 5                                                                                & 2           & 0           & 3           & 4            & 0            & 1           & 4           & 3           & 1           & 3           & \multicolumn{1}{c|}{0}          & \multicolumn{1}{c|}{2}           \\ \hline
manifoldcf                                                                          & 5                                                                                & 3           & 0           & 2           & 3            & 0            & 2           & 3           & 2           & 2           & 3           & \multicolumn{1}{c|}{0}          & \multicolumn{1}{c|}{2}           \\ \hline
meecrowave                                                                          & 3                                                                                & 3           & 0           & 0           & 2            & 0            & 1           & 2           & 0           & 1           & 2           & \multicolumn{1}{c|}{0}          & \multicolumn{1}{c|}{1}           \\ \hline
spark                                                                               & 27                                                                               & 23          & 0           & 4           & 27           & 0            & 0           & ---         & ---         & ---         & 4           & \multicolumn{1}{c|}{0}          & \multicolumn{1}{c|}{23}          \\ \hline
tika                                                                                & 0                                                                                & 0           & 0           & 0           & 0            & 0            & 0           & 0           & 3           & 0           & 0           & \multicolumn{1}{c|}{0}          & \multicolumn{1}{c|}{0}           \\ \hline
tomee                                                                               & 9                                                                                & 6           & 0           & 3           & 6            & 0            & 3           & 4           & 2           & 5           & 4           & \multicolumn{1}{c|}{0}          & \multicolumn{1}{c|}{5}           \\ \hline
wicket                                                                              & 5                                                                                & 2           & 0           & 3           & 5            & 0            & 0           & 2           & 3           & 3           & 0           & \multicolumn{1}{c|}{0}          & \multicolumn{1}{c|}{5}           \\ \hline
artemis-commons                                                                     & 15                                                                               & 13          & 0           & 2           & 13           & 0            & 2           & ---         & ---         & ---         & ---         & \multicolumn{1}{c|}{---}        & \multicolumn{1}{c|}{---}         \\ \hline 
\multicolumn{1}{|c|}{\textbf{Total}}                                                & \textbf{88}                                                                      & \textbf{63} & \textbf{0}  & \textbf{25} & \textbf{67}  & \textbf{0}   & \textbf{21} & \textbf{35} & \textbf{33} & \textbf{11} & \textbf{23} & \multicolumn{1}{c|}{\textbf{0}} & \multicolumn{1}{c|}{\textbf{50}} \\ \hline
\end{tabular}
\vspace{-4mm}
\end{table*}

\begin{itemize}
    \item
    In the prospect of path sensitivity, it is obvious that none of the cryptographic vulnerability detection tools is path-sensitive in their static analysis. The tools generate 10, 13, 13, 12 false positive alerts for path sensitive cases, respectively. The possible reason for the false positive alert is that for the concerned variable, a container is defined to store all values of the concerned variable. There is no ordered list that shows the latest assignment. Therefore, alerts will be raised if the container contains any vulnerable value that is intended to be used in the Crypto API. A significant reason for having a high false positive rate because of the tools being path insensitive.
    \item
    SpotBugs is not designed to capture vulnerability threats in advanced cases. Therefore, it shows 0\% precision and recall. 
    
    \item
    SpotBugs produces 12 false positives for combined cases. In combined cases, SpotBugs failed to detect the source of vulnerability using both interprocedural and field sensitive analysis. For example, in Symmetric Cipher cases, instead of showing the correct ``CIPHER\_INTEGRITY" alert, it produces an incorrect ``HARD\_CODE\_PASSWORD" alert.
    
    \item CryptoGuard performs better than other tools in terms of both precision and recall. The reason behind this is 1) Cryptoguard performs dataflow analysis based on forward slicing and backward slicing that efficiently handles the advanced cases, 2) CryptoGuard follows several refinement insights that systematically remove irrelevant constants, hence reduced false positives.  However, as being a static analysis tool, CryptoGuard cannot handle path-sensitive cases. In addition, CryptoGuard missed 3 vulnerabilities due to clipping orthogonal method invocation (i.e., limiting the depth to visit callee method).  
    

    \item CrySL produces incorrect ``RequiredPredicateError" alerts for the cryptographic key, password in PBE, password in KeyStore misuse test cases that contribute to generate a high false positive rate. The reason is that the cryptographic APIs used in these cases follow strict rules in CrySL. Therefore, even if we use a secure unpredictable byte array as an argument for crypto APIs, it still generates incorrect alerts.

    
    \item Coverity is not designed to detect vulnerable ciphers and cryptographic hash functions in advanced cases. That is the reason for having high false negative values and generating high FNR in Coverity. Coverity is a closed sourced detection tool. Therefore, we are unable to confirm the reason for the incorrect detection cases.
    

\end{itemize}

\vspace{-1mm}

In summary, for all of the advanced cases, SpotBugs is not designed to identify the advanced vulnerability threats correctly. Therefore, the precision rate is 0\%. CryptoGuard detects fairly well (missed only 3 cases) among all detection tools with a precision of 83.33\%. For CrySL produce precision of 56.34\%. Coverity generates a precision of 52.00\%.

\vspace{-3mm}
\subsection{\apachebenchname{}: Analysis of Results}

TABLE~\ref{tab:apache_result} presents the number of true positive and false positive vulnerability threats detected by the tools. CrySL fails to analyze spark and artemis-commons project. Coverity fails to analyze artemis-commons project. SpotBugs and CryptoGuard successfully analyze all 10 projects. Overall, we capture the following findings.

\begin{itemize}
    \item  SpotBugs, CryptoGuard, and Coverity do not generate any false positive. CrySL has a high false positive value of 33. We have already discussed the reason for generating high false positive for CrySL in Section 6.3.2. 
    
    \item SpotBugs, CryptoGuard, CrySL, and Coverity can accurately detect 63, 67, and 35 and 23 alerts respectively from 88 alerts. The main reason for missed alarms is that no tool can detect all 18 types of vulnerabilities as shown in TABLE~\ref{table:crypto_alert_keywords}. For example, SpotBugs and CryptoGuard cannot capture vulnerable crypto algorithm usage in SecretKeySpec API. Among the successfully compiled programs (i.e., from 8 Apache projects), CrySL capture 33 out of 44.     

    \item After analyzing ten Apache projects, we find that there are 79 basic cases, whereas, the number of advance cases is only 43. Therefore, in the real world codes, the number of basic cases is much higher than advanced cases. Vulnerability detection tools should consider expanding their coverage to detect more categories of vulnerabilities.
    
    \item From TABLE~\ref{tab:apache_result}, we observe that  CrySL fails to analyze two Apache projects: spark and artemis-commons. CrySL throws StackOverFlowError (i.e., memory error) during analyzing objects for spark. The probable reason is the larger number of files and lines of code Spark contains for analysis. For artemis-commons, CrySL throws NullPointerErrorException during analysis due to the reference variable not pointing to any object. Coverity fails to analyze only the artemis-commons project. Coverity is closed source, therefore, we are unable to confirm the reason for this failure. TABLE~\ref{tab:apache_runtime} shows the runtime on Apache projects for only CryptoGuard and CrySL. For Coverity and SpotBugs, we use the web version that takes all scan requests for users and reports results after complete scanning. Therefore, we cannot calculate their original runtime for comparison. Among the 8 successful analyzed projects, we observe average runtime for CrySL is 14.64 seconds and CryptoGuard is 11.46 seconds. For the largest Apache project Spark (LoC: 311,856), CryptoGuard successfully analyzes in 88.68 seconds and CrySL shows the failure of analysis report after 46.84 seconds. Overall, SpotBugs and CryptoGuard successfully analyze all 10 Apache projects. Therefore, SpotBugs, CryptoGuard are scalable for large projects.

\end{itemize}

\begin{table}[]
\centering
\caption{Runtime for analyzing Apache projects. Star (*) symbol indicates that the analysis was unsuccessful.}
\vspace{-3mm}
\label{tab:apache_runtime}
\begin{tabular}{|l|c|c|c|}
\hline
\rowcolor[HTML]{C0C0C0} 
\cellcolor[HTML]{C0C0C0}   & \cellcolor[HTML]{C0C0C0}                                          & \multicolumn{2}{c|}{\cellcolor[HTML]{C0C0C0}\textbf{Runtime (sec)}} \\ \cline{3-4} 
\rowcolor[HTML]{C0C0C0} 
\multirow{-2}{*}{\cellcolor[HTML]{C0C0C0}\textbf{Apache Projects}} & \multirow{-2}{*}{\cellcolor[HTML]{C0C0C0}\textbf{LoC}} & \textbf{CryptoGuard}                & \textbf{CrySL}                \\ \hline
deltaspike      &  5.1K                                                                    & 4.31                    & 6.95              \\ \hline
directory-server  &  20.8K                                                                  & 8.96                    & 23.03             \\ \hline
incubator-taverna-workbench &  9.9K                                                        & 12.69                   & 7.94              \\ \hline
manifoldcf      &  17K                                                                    & 7.07                    & 8.20              \\ \hline
meecrowave         &  5.6K                                                                 & 4.67                    & 7.24              \\ \hline
spark                 &  311.9K                                                             & 88.68                   & 46.84*            \\ \hline
tika        &  16.6K                                                                        & 7.46                    & 8.15              \\ \hline
tomee          &  118.7K                                                                     & 40.52                   & 34.81             \\ \hline
wicket            &  13.4K                                                                  & 5.99                    & 20.83             \\ \hline
artemis-commons      &  8.9K                                                               & 5.63                    & 19.82*            \\ \hline
\end{tabular}
\vspace{-4mm}
\end{table}

\vspace{-4mm}
\subsection{Verifiability}
Our benchmarks are open-sourced and are available on GitHub~\cite{cryptoapibench-github,apachecryptoapibench-github}. It contains the Java cryptographic API test cases. The detailed documentation and explanation are provided there.

\vspace{-4mm}
\section{Discussion}
\label{sec:discussion}

\noindent
{\bf Tool insights.} No tool can cover all categories of vulnerabilities (TABLE~\ref{table:totalbench}). However, their methodologies can be extended to cover most of these vulnerabilities. For example, the technique that Coverity uses to detect constant cryptographic keys can be transferred to detect static IVs or fewer iteration counts.

The main differences among different tools are within their approach to trade-offs among false positives, false negatives. Our experimental evaluation reveals that all of these tools produce a number of false positives and false negatives. CryptoGuard performs on-demand inter-procedural dataflow analysis. Its backward data flow analysis starts from the slicing criteria and explores upward ($\uparrow$) and orthogonally ($\rightarrow$) on-demand. Orthogonal method invocation chains always return to the call sites. By leveraging this insight, CryptoGuard offers a performance vs scalability tradeoff by limiting the depth of the orthogonal invocations (which is “clipping of orthogonal method invocations"). In the current implementation, the depth is set to be 1. That means CryptoGuard will skip deeper orthogonal callee methods, which may result in false negatives. However, the advantage of the orthogonal method invocation technique is that it helps to improve precision. 


The main focus of CrySL is to provide a language to specify a class of cryptographic misuse vulnerabilities that can be detected using a generic detection engine. A prime reason behind the false positives can be the strictness of the rule definitions that is inherited from the language itself. For example, CrySL raises an alert if a cryptographic key is not generated using a key generator. However, one can legitimately reuse a previously generated key, which CrySL would mistakenly detect as a vulnerability. An impressive aspect of CrySL is that it is constantly being maintained and updated to improve its accuracy. The methodology of SpotBugs is inherently limited to detect advanced cases as they use patterns to detect most of the vulnerabilities.

None of these tools are path-sensitive, i.e., all raising false alerts in path sensitive cases. A possible reason for this failure is that the existing path-sensitive analysis techniques are usually costly, i.e., high runtime complexity.

CryptoAPI-Bench cannot be used to evaluate scalability property. All of our test cases are lightweight by design. The primary focus is to produce easily readable test cases that demand minimal code to express complex program properties. On the other hand, all of the projects in \apachebenchname{} are complex programs including a lot of files and lines of code. The primary focus is to test the vulnerability detection tool's scalability property and extrapolation to applications on real-world code.

{\bf Our limitation.} Currently, our benchmark does not contain cryptographic cases, e.g., digital signature, CBC-MAC misuses in MAC, other modes of operations (e.g., CTR). We plan to include test cases based on these cryptographic vulnerabilities in our CryptoAPI-Bench benchmark. Furthermore, our benchmark does not have any cases that involve Java reflection APIs. The primary reason is that the use of Java reflection during cryptographic coding is highly unlikely. Consequently, none of the existing open-sourced tools is designed to detect such cases. However, we plan to include new cases that leverage Java reflection APIs to induce cryptographic misuse vulnerabilities.
\vspace{-3mm}
\section{Related Work}
\label{sec:related_work}
{\bf Vulnerability detection benchmarks.} 
AndroZoo++~\cite{DBLP:journals/corr/abs-1709-05281} is a collection of over eight million Android apps~\cite{androzooweb} that drives a lot of security, software engineering, and malware analysis research. However, vulnerabilities in these apps are not documented, hence not suitable for vulnerability detection benchmarking purposes.

DroidBench~\cite{DBLP:conf/pldi/ArztRFBBKTOM14}, a benchmark containing vulnerable android apps, fills the gap by providing specific vulnerability locations within the benchmark. Till date, DroidBench is one of the most popular benchmarks to evaluate the performance of vulnerability detection tools in Android literature. In total, DroidBench has 119 APKs from 13 categories (Commit id 0fe281b). Categories include vulnerabilities that use field and object sensitivity, inter-app communication, inter-component communication, android life-cycle, reflection, etc. However, DroidBench {\em i)} does not cover cryptographic misuse vulnerabilities and {\em ii)} does not have source code. To the best of our knowledge, Ghera~\cite{DBLP:conf/promise/MitraR17} is the only Android app benchmark that contains app source code. Like DroidBench, most of the vulnerabilities in Ghera are specific to Android apps and barely contain any cryptographic misuse vulnerabilities. To be specific, CryptoAPI-Bench and Ghera have only 2 types of vulnerabilities in common.

OWASP Benchmark~\cite{burato2017security} is fundamentally designed to capture eleven cybersecurity vulnerabilities. However, among the detected vulnerabilities, it builds to address only three Java cryptographic vulnerabilities, i.e., weak encryption algorithm, weak hash algorithm, and a weak random number.   

SonarSource~\cite{sonarsource} released a set of vulnerability samples that can be useful to check for coverage of vulnerability categories. A verification tool for five common audit controls is proposed for ensuring continuous compliance~\cite{kellogg2020continuous}.

{\bf Other benchmarks.} The DaCapo benchmarks~\cite{DBLP:conf/oopsla/BlackburnGHKMBDFFGHHJLMPSVDW06} are designed to evaluate the performance of various components of Java virtual machine (JVM), Garbage collection (GC), Just-in-time (JIT) compiler itself.  BugBench~\cite{lu2005bugbench} is a benchmark to find C/C++ bugs that contains 17 real-world applications. BugBench mostly covers various memory, concurrency, and semantic bugs. To detect bugs in the multi-threaded Java programs, a benchmark and framework have been proposed~\cite{havelund2003benchmark,eytani2007towards}. For dynamic software updating system, a standardized benchmark system is proposed to check the system's practicality, flexibility, and usability~\cite{smith2012towards}. Coding practice and recommendations are provided for 28 enterprise applications that use Spring security framework~\cite{islam2020coding}. ManyBugs and IntroClass benchmarks are designed to evaluate various C/C++ code repair techniques~\cite{DBLP:journals/tse/GouesHSBDFW15}. Most of the defects in ManyBugs and IntroClass do not impact security, e.g., in the ManyBugs benchmark, more than half of the instances impact correctness, not necessary security.


\vspace{-3mm}
\section{Conclusion and Future Work}
\label{conclusion}

We believe that for scientific, in-depth, and reproducible comparisons benchmark is an important component. In this paper, we present CryptoAPI-Bench and \apachebenchname{} to evaluate the detection accuracy, scalability, and security guarantees of various cryptographic misuse detection tools. Our benchmarks are open-sourced and are available on GitHub. We evaluated four static analysis tools that are capable of detecting cryptographic misuses. Our evaluation revealed some interesting insights, i.e., {\em i)} tools that are specialized to detect cryptographic misuses (e.g., CryptoGuard, CrySL) cover more rules and higher recall than general purpose tools (e.g., SpotBugs, Coverity), {\em ii)} none of the existing tools is path-sensitive. 

We are actively working on expanding CryptoAPI-Bench by adding new rules, test cases, and covering new cryptographic APIs. In the future, we plan to achieve the following goals.

\begin{itemize}
     \item To motivate the research of cryptographic misuse detection tools for other platforms, we plan to extend CryptoAPI-Bench to cover other popular languages, e.g., Python.
    
    \item Other non-cryptographic API misuses (e.g., Android APIs to access sensitive information (location, IMEI, passwords, etc.)~\cite{DBLP:conf/ccs/BosuLYW17, DBLP:conf/ndss/NanY0ZZ018}, fingerprint protection~\cite{DBLP:conf/ndss/BianchiFMKVCL18}, cloud service APIs for information storage~\cite{zuodoes}) are also proven to cause catastrophic security consequences. We also plan to include the misuses of these critical non-cryptographic APIs.
\end{itemize}


%



\vspace{-5mm}
\ifCLASSOPTIONcompsoc
  \section*{Acknowledgments}
\else
  \section*{Acknowledgment}
\fi
This work has been supported by the National Science Foundation under Grant No. CNS-1929701 and the Virginia  Commonwealth Cyber Initiative (CCI).

\ifCLASSOPTIONcaptionsoff
  \newpage
\fi



%



\vspace{-5mm}
\bibliographystyle{IEEEtran}
\bibliography{paper}

\begin{thebibliography}{10}
\providecommand{\url}[1]{#1}
\csname url@samestyle\endcsname
\providecommand{\newblock}{\relax}
\providecommand{\bibinfo}[2]{#2}
\providecommand{\BIBentrySTDinterwordspacing}{\spaceskip=0pt\relax}
\providecommand{\BIBentryALTinterwordstretchfactor}{4}
\providecommand{\BIBentryALTinterwordspacing}{\spaceskip=\fontdimen2\font plus
\BIBentryALTinterwordstretchfactor\fontdimen3\font minus
  \fontdimen4\font\relax}
\providecommand{\BIBforeignlanguage}[2]{{%
\expandafter\ifx\csname l@#1\endcsname\relax
\typeout{** WARNING: IEEEtran.bst: No hyphenation pattern has been}%
\typeout{** loaded for the language `#1'. Using the pattern for}%
\typeout{** the default language instead.}%
\else
\language=\csname l@#1\endcsname
\fi
#2}}
\providecommand{\BIBdecl}{\relax}
\BIBdecl

\bibitem{DBLP:conf/ccs/FahlHMSBF12}
S.~Fahl, M.~Harbach, T.~Muders \emph{et~al.}, ``Why {E}ve and {M}allory {L}ove
  {A}ndroid: {A}n {A}nalysis of {A}ndroid {SSL} (in) {S}ecurity,'' in \emph{the
  {ACM} Conference on Computer and Communications Security, CCS'12, Raleigh,
  NC, USA, October 16-18, 2012}, 2012, pp. 50--61.

\bibitem{DBLP:conf/ccs/GeorgievIJABS12}
M.~Georgiev, S.~Iyengar, S.~Jana \emph{et~al.}, ``The {M}ost {D}angerous {C}ode
  in the {W}orld: {V}alidating {SSL} {C}ertificates in {N}on-{B}rowser
  {S}oftware,'' in \emph{the {ACM} Conference on Computer and Communications
  Security, CCS'12, Raleigh, NC, USA, October 16-18}, 2012, pp. 38--49.

\bibitem{DBLP:conf/ccs/EgeleBFK13}
M.~Egele, D.~Brumley, Y.~Fratantonio \emph{et~al.}, ``An {E}mpirical {S}tudy of
  {C}ryptographic {M}isuse in {A}ndroid {A}pplications,'' in \emph{ACM
  Conference on Computer and Communications Security, {CCS}'13}, 2013, pp.
  73--84.

\bibitem{Meng-ICSE-2018}
N.~Meng, S.~Nagy, D.~Yao \emph{et~al.}, ``Secure {C}oding {P}ractices in
  {J}ava: {C}hallenges and {V}ulnerabilities,'' in \emph{International
  Conference on Software Engineering, {ICSE}'18}, Gothenburg, Sweden, May 2018.

\bibitem{DBLP:journals/corr/abs-1806-06881}
S.~Rahaman, Y.~Xiao, S.~Afrose \emph{et~al.}, ``{CryptoGuard}: {H}igh
  {P}recision {D}etection of {C}ryptographic {V}ulnerabilities in
  {M}assive-sized {J}ava {P}rojects,'' in \emph{ACM Conference on Computer and
  communications security, CCS'19}, London, UK, Nov. 2019.

\bibitem{DBLP:conf/secdev/RahamanY17}
\BIBentryALTinterwordspacing
S.~Rahaman and D.~Yao, ``Program {A}nalysis of {C}ryptographic
  {I}mplementations for {S}ecurity,'' in \emph{{IEEE} Cybersecurity
  Development, SecDev'17, Cambridge, MA, USA, September 24-26}, 2017, pp.
  61--68. [Online]. Available: \url{https://doi.org/10.1109/SecDev.2017.23}
\BIBentrySTDinterwordspacing

\bibitem{SP-crypto-API-2017}
Y.~Acar, M.~Backes, S.~Fahl \emph{et~al.}, ``Comparing the {U}sability of
  {C}ryptographic {API}s,'' in \emph{{IEEE} Symposium on Security and Privacy,
  {SP'17}, San Jose, CA, USA, May 22-26}, 2017, pp. 154--171.

\bibitem{Bodden-crypto-API-2015}
S.~Nadi, S.~Kr\"{u}ger, M.~Mezini \emph{et~al.}, ``Jumping {T}hrough {H}oops:
  {W}hy {D}o {J}ava {D}evelopers {S}truggle with {C}ryptography {API}s?'' in
  \emph{International Conference on Software Engineering, {ICSE}'16}, 2016, pp.
  935--946.

\bibitem{DBLP:conf/sp/OltroggeDSAFRPB18}
M.~Oltrogge, E.~Derr, C.~Stransky \emph{et~al.}, ``The {R}ise of the {C}itizen
  {D}eveloper: {A}ssessing the {S}ecurity {I}mpact of {O}nline {A}pp
  {G}enerators,'' in \emph{{IEEE} Symposium on Security and Privacy, {SP'18},
  San Francisco, California, {USA}, 21-23 May}, 2018, pp. 634--647.

\bibitem{DBLP:conf/sp/AcarBFKMS16}
Y.~Acar, M.~Backes, S.~Fahl \emph{et~al.}, ``You {G}et {W}here {Y}ou're
  {L}ooking for: {T}he {I}mpact of {I}nformation {S}ources on {C}ode
  {S}ecurity,'' in \emph{{IEEE} Symposium on Security and Privacy, {SP'16}, San
  Jose, CA, USA, May 23-25}, 2016, pp. 289--305.

\bibitem{DBLP:conf/soups/AssalC18}
H.~Assal and S.~Chiasson, ``Security in the {S}oftware {D}evelopment
  {L}ifecycle,'' in \emph{Fourteenth Symposium on Usable Privacy and Security,
  {SOUPS'18}, Baltimore, MD, USA, August 12-14}, 2018, pp. 281--296.

\bibitem{DBLP:conf/kbse/KrugerNRAMBGGWD17}
S.~Kr{\"{u}}ger \emph{et~al.}, ``{CogniCrypt}: {S}upporting {D}evelopers in
  using {C}ryptography,'' in \emph{IEEE/ACM International Conference on
  Automated Software Engineering, {ASE}'17}, 2017, pp. 931--936.

\bibitem{DBLP:conf/ecoop/KrugerS0BM18}
S.~Kr{\"{u}}ger, J.~Sp{\"{a}}th, K.~Ali \emph{et~al.}, ``{CrySL}: An
  {E}xtensible {A}pproach to {V}alidating the {C}orrect {U}sage of
  {C}ryptographic {API}s,'' in \emph{European Conference on Object-Oriented
  Programming, {ECOOP}'18}, 2018, pp. 10:1--10:27.

\bibitem{DBLP:conf/ccs/NguyenWABWF17}
D.~C. Nguyen \emph{et~al.}, ``A {S}titch in {T}ime: {S}upporting {A}ndroid
  {D}evelopers in {W}riting {S}ecure {C}ode,'' in \emph{ACM Conference on
  Computer and Communications Security, {CCS}'17}, 2017, pp. 1065--1077.

\bibitem{piccolboni2020crylogger}
L.~Piccolboni, G.~Di~Guglielmo, L.~P. Carloni \emph{et~al.}, ``Crylogger:
  {D}etecting {C}rypto {M}isuses {D}ynamically,'' \emph{arXiv preprint
  arXiv:2007.01061}, 2020.

\bibitem{DBLP:conf/ndss/SounthirarajSGLK14}
D.~Sounthiraraj, J.~Sahs, G.~Greenwood \emph{et~al.}, ``{SMV}-{H}unter: {L}arge
  {S}cale, {A}utomated {D}etection of {SSL/TLS} {M}an-in-the-{M}iddle
  {V}ulnerabilities in {A}ndroid {A}pps,'' in \emph{The Network and Distributed
  System Security Symposium, {NDSS}'14}, 2014.

\bibitem{DBLP:conf/fps/GagnonFFDOB15}
F.~Gagnon, M.~Ferland, M.~Fortier \emph{et~al.}, ``{AndroSSL}: {A} {P}latform
  to {T}est {A}ndroid {A}pplications {C}onnection {S}ecurity,'' in
  \emph{International Symposium on Foundations and Practice of Security,
  {FPS}'15}, 2015, pp. 294--302.

\bibitem{DBLP:conf/sp/AshcraftE02}
K.~Ashcraft and D.~R. Engler, ``Using {P}rogrammer-{W}ritten {C}ompiler
  {E}xtensions to {C}atch {S}ecurity {H}oles,'' in \emph{{IEEE} Symposium on
  Security and Privacy, (SP'02), Berkeley, California, USA, May 12-15}, 2002,
  pp. 143--159.

\bibitem{DBLP:conf/uss/MachirySCSKV17}
A.~Machiry, C.~Spensky, J.~Corina \emph{et~al.}, ``{DR.} {CHECKER:} {A}
  {S}oundy {A}nalysis for {L}inux {K}ernel {D}rivers,'' in \emph{26th {USENIX}
  Security Symposium, {USENIX} Security'17, Vancouver, BC, Canada, August
  16-18, 2017}, 2017, pp. 1007--1024.

\bibitem{wiki:dataflow}
\BIBentryALTinterwordspacing
{Wikipedia contributors}, ``Data-flow {A}nalysis --- {W}ikipedia{,} {T}he
  {F}ree {E}ncyclopedia,'' last accessed: September 9, 2021. [Online].
  Available: \url{https://en.wikipedia.org/wiki/Data-flow\_analysis}
\BIBentrySTDinterwordspacing

\bibitem{10.1145/3290361}
\BIBentryALTinterwordspacing
J.~Sp\"{a}th, K.~Ali, and E.~Bodden, ``Context-, {F}low-, and {F}ield-sensitive
  {D}ata-flow {A}nalysis {U}sing {S}ynchronized {P}ushdown {S}ystems,''
  \emph{Proc. ACM Program. Lang.}, vol.~3, no. POPL, Jan. 2019. [Online].
  Available: \url{https://doi.org/10.1145/3290361}
\BIBentrySTDinterwordspacing

\bibitem{DBLP:conf/pldi/ArztRFBBKTOM14}
S.~Arzt, S.~Rasthofer, C.~Fritz \emph{et~al.}, ``{FlowDroid}: {P}recise
  {C}ontext, {F}low, {F}ield, {O}bject-{S}ensitive and {L}ifecycle-aware
  {T}aint {A}nalysis for {A}ndroid {A}pps,'' in \emph{{ACM} {SIGPLAN}
  Conference on Programming Language Design and Implementation, {PLDI}'14},
  2014, pp. 259--269.

\bibitem{DBLP:conf/promise/MitraR17}
J.~Mitra and V.~Ranganath, ``Ghera: {A} {R}epository of {A}ndroid {A}pp
  {V}ulnerability {B}enchmarks,'' in \emph{Proceedings of the 13th
  International Conference on Predictive Models and Data Analytics in Software
  Engineering, {PROMISE}'17, Toronto, Canada, November 8, 2017}, 2017, pp.
  43--52.

\bibitem{spotbugs}
``Spot{B}ugs: {F}ind {B}ugs in {J}ava {P}rograms,''
  https://spotbugs.github.io/, online; Last accessed: December 3, 2020.

\bibitem{coverity}
``Coverity {S}tatic {A}pplication {S}ecurity {T}esting ({SAST}),''
  https://www.synopsys.com/software-integrity/security-testing/\\static-analysis-sast.html,
  online; Last accessed: December 3, 2020.

\bibitem{cryptoapibench-github}
S.~Afrose, ``Crypto{API}-{B}ench,''
  \url{https://github.com/CryptoAPI-Bench/CryptoAPI-Bench}, 2019.

\bibitem{apachecryptoapibench-github}
------, ``Apache{C}rypto{API}-{B}ench,''
  \url{https://github.com/CryptoAPI-Bench/ApacheCryptoAPI-Bench}, 2020.

\bibitem{afrose2019cryptoapi}
S.~Afrose, S.~Rahaman, and D.~Yao, ``Crypto{API}-{B}ench: {A} {C}omprehensive
  {B}enchmark on {J}ava {C}ryptographic {API} {M}isuses,'' in \emph{2019 IEEE
  Cybersecurity Development (SecDev)}.\hskip 1em plus 0.5em minus 0.4em\relax
  IEEE, 2019, pp. 49--61.

\bibitem{nistdoc}
E.~Barker and A.~Roginsky, ``\BIBforeignlanguage{en}{Transitioning the {U}se of
  {C}ryptographic {A}lgorithms and {K}ey {L}engths},'' in
  \emph{\BIBforeignlanguage{en}{Special Publication (NIST SP), National
  Institute of Standards and Technology, Gaithersburg, MD}}, 2019.

\bibitem{244246}
E.~Barker, L.~Chen, A.~Roginsky \emph{et~al.},
  ``\BIBforeignlanguage{en}{Recommendation for {P}air-wise {K}ey
  {E}stablishment {U}sing {I}nteger {F}actorization {C}ryptography},'' in
  \emph{\BIBforeignlanguage{en}{Special Publication (NIST SP), National
  Institute of Standards and Technology, Gaithersburg, MD}}, 2019.

\bibitem{nistcsrc}
``{NIST} {C}omputer {S}ecurity {R}esource {C}enter,''
  \url{https://csrc.nist.gov/projects}, online; Last accessed: September 3,
  2021.

\bibitem{sonarsource}
``{S}onar{S}ource {S}tatic {C}ode {A}nalysis,'' https://rules.sonarsource.com/,
  online; Last accessed: December 3, 2020.

\bibitem{UrlSpoofing}
``{URL} {S}poofing,''
  \url{http://www.securitysupervisor.com/security-q-a/network-security/262-what-is-url-spoofing.html},
  online; Last accessed: December 3, 2020.

\bibitem{BugsPatternTrustManager}
``Find {S}ecurity {B}ugs,'' \url{https://find-sec-bugs.github.io/}, online;
  Last accessed: December 3, 2020.

\bibitem{HostnameVerificationSSLSocket}
``Hostname {V}erification to {SSL} {S}ocket,''
  \url{https://www.thecodingforums.com/threads/adding-hostname-verification-to-sslsocket.958287/},
  online; Last accessed: December 3, 2020.

\bibitem{barton2006transferring}
C.~A. Barton, G.~A. Clarke, and S.~Crowe, ``Transferring {D}ata via a {S}ecure
  {N}etwork {C}onnection,'' 2006, {US} Patent 7,093,121.

\bibitem{knuth2014art}
D.~E. Knuth, \emph{Art of {C}omputer {P}rogramming, volume 2: {S}eminumerical
  {A}lgorithms}.\hskip 1em plus 0.5em minus 0.4em\relax Addison-Wesley
  Professional, 2014.

\bibitem{moriarty2017pkcs}
K.~Moriarty, B.~Kaliski, and A.~Rusch, ``{PKCS} \#5: {P}assword-{B}ased
  {C}ryptography {S}pecification {V}ersion 2.1,'' 2017.

\bibitem{AESEncryption}
``{AES} {E}ncryption,'' \url{https://aesencryption.net/}, online; Last
  accessed: December 3, 2020.

\bibitem{bellare2015new}
M.~Bellare, ``New {P}roofs for {NMAC} and {HMAC}: {S}ecurity {W}ithout
  {C}ollision {R}esistance,'' \emph{Journal of Cryptology}, vol.~28, no.~4, pp.
  844--878, 2015.

\bibitem{stringimmutable}
``Oracle,''
  \url{https://docs.oracle.com/javase/8/docs/api/java/secur-ity/Permission.html},
  online; Last accessed: September 3, 2021.

\bibitem{secureCode}
``Secure {C}ode {R}eview: 8 {S}ecurity {C}ode {R}eview {B}est {P}ractices,''
  \url{https://snyk.io/blog/secure-code-review/}, online; Last accessed:
  September 3, 2021.

\bibitem{passwordmutable}
``Secure {C}oding {G}uidelines for {J}ava {SE},''
  \url{https://www.oracle.com/java/technologies/javase/seccodeguide.html},
  online; Last accessed: September 3, 2021.

\bibitem{grammatech}
``{GRAMMATECH},'' https://www.grammatech.com/, online; Last accessed: December
  3, 2020.

\bibitem{qark}
``{Quick {A}ndroid {R}eview {K}it (QARK)},'' https://github.com/linkedin/qark,
  online; Last accessed: December 3, 2020.

\bibitem{swamp}
``Welcome to the {SWAMP},'' \url{https://continuousassurance.org}, 2018.

\bibitem{transitionswamp}
``Transition of {SWAMP} software,'' https://continuousassurance.org/blog/,
  online; Last accessed:December 3, 2020.

\bibitem{cryptoguardgithub}
``Crypto{G}uard,'' https://github.com/CryptoGuardOSS/cryptoguard, online; Last
  accessed: December 3, 2020.

\bibitem{cryslgithub}
``Cognicrypt\_{SAST}: {C}ry{SL}\-to\-{S}tatic {A}nalysis {C}ompiler,''
  https://github.com/CROSSINGTUD/CryptoAnalysis, online; Last accessed:
  December 3, 2020.

\bibitem{DBLP:journals/corr/abs-1709-05281}
L.~Li, J.~Gao, M.~Hurier \emph{et~al.}, ``Andro{Z}oo++: {C}ollecting {M}illions
  of {A}ndroid {A}pps and {T}heir {M}etadata for the {R}esearch {C}ommunity,''
  \emph{arXiv preprint arXiv:1709.05281}, 2017.

\bibitem{androzooweb}
``Andro{Z}oo,'' https://androzoo.uni.lu/, online; Last accessed: December 3,
  2020.

\bibitem{burato2017security}
E.~Burato, P.~Ferrara, and F.~Spoto, ``Security {A}nalysis of the {OWASP}
  {B}enchmark with {J}ulia,'' \emph{The Italian Conference on CyberSecurity
  (ITASEC)}, vol.~17, 2017.

\bibitem{kellogg2020continuous}
M.~Kellogg, M.~Sch{\"a}f, S.~Tasirans \emph{et~al.}, ``Continuous
  {C}ompliance,'' in \emph{35th IEEE/ACM International Conference on Automated
  Software Engineering (ASE)}, 2020, pp. 511--523.

\bibitem{DBLP:conf/oopsla/BlackburnGHKMBDFFGHHJLMPSVDW06}
S.~M. Blackburn \emph{et~al.}, ``The {D}a{C}apo {B}enchmarks: {J}ava
  {B}enchmarking {D}evelopment and {A}nalysis,'' in \emph{Proceedings of the
  21th Annual {ACM} {SIGPLAN} Conference on Object-Oriented Programming,
  Systems, Languages, and Applications, {OOPSLA'06}, Portland, Oregon, {USA},
  October 22-26}, 2006, pp. 169--190.

\bibitem{lu2005bugbench}
S.~Lu, Z.~Li, F.~Qin \emph{et~al.}, ``Bug{B}ench: {B}enchmarks for {E}valuating
  {B}ug {D}etection {T}ools,'' in \emph{Workshop on the Evaluation of Software
  Defect Detection Tools}, vol.~5, 2005.

\bibitem{havelund2003benchmark}
K.~Havelund, S.~D. Stoller, and S.~Ur, ``Benchmark and {F}ramework for
  {E}ncouraging {R}esearch on {M}ulti-{T}hreaded {T}esting {T}ools,'' in
  \emph{Proceedings International Parallel and Distributed Processing
  Symposium, IPDPS'03}.\hskip 1em plus 0.5em minus 0.4em\relax IEEE, 2003, p.
  286.

\bibitem{eytani2007towards}
Y.~Eytani, K.~Havelund, S.~D. Stoller \emph{et~al.}, ``Towards a {F}ramework
  and a {B}enchmark for {T}esting {T}ools for {M}ulti-{T}hreaded {P}rograms,''
  \emph{Concurrency and Computation: Practice and Experience}, vol.~19, no.~3,
  pp. 267--279, 2007.

\bibitem{smith2012towards}
E.~K. Smith, M.~Hicks, and J.~S. Foster, ``Towards {S}tandardized {B}enchmarks
  for {D}ynamic {S}oftware {U}pdating {S}ystems,'' in \emph{4th International
  Workshop on Hot Topics in Software Upgrades (HotSWUp'12)}.\hskip 1em plus
  0.5em minus 0.4em\relax IEEE, 2012, pp. 11--15.

\bibitem{islam2020coding}
M.~Islam, S.~Rahaman, N.~Meng \emph{et~al.}, ``Coding {P}ractices and
  {R}ecommendations of {S}pring {S}ecurity for {E}nterprise {A}pplications,''
  in \emph{2020 IEEE Secure Development (SecDev)}.\hskip 1em plus 0.5em minus
  0.4em\relax IEEE, 2020, pp. 49--57.

\bibitem{DBLP:journals/tse/GouesHSBDFW15}
C.~{Le Goues}, N.~Holtschulte, E.~K. Smith \emph{et~al.}, ``{T}he {M}any{B}ugs
  and {I}ntro{C}lass {B}enchmarks for {A}utomated {R}epair of {C} {P}rograms,''
  \emph{IEEE Transactions on Software Engineering}, vol.~41, no.~12, pp.
  1236--1256, 2015.

\bibitem{DBLP:conf/ccs/BosuLYW17}
A.~Bosu, F.~Liu, D.~Yao \emph{et~al.}, ``Collusive {D}ata {L}eak and {M}ore:
  {L}arge-{S}cale {T}hreat {A}nalysis of {I}nter-app {C}ommunications,'' in
  \emph{ACM ASIA Conference on Computer and Communications Security,
  {AsiaCCS'17}}, 2017, pp. 71--85.

\bibitem{DBLP:conf/ndss/NanY0ZZ018}
Y.~Nan, Z.~Yang, X.~Wang \emph{et~al.}, ``Finding {C}lues for {Y}our {S}ecrets:
  {S}emantics-{D}riven, {L}earning-{B}ased {P}rivacy {D}iscovery in {M}obile
  {A}pps,'' in \emph{25th Annual Network and Distributed System Security
  Symposium, {NDSS'18}, San Diego, California, USA, February 18-21}, 2018.

\bibitem{DBLP:conf/ndss/BianchiFMKVCL18}
A.~Bianchi, Y.~Fratantonio, A.~Machiry \emph{et~al.}, ``Broken {F}ingers: On
  the {U}sage of the {F}ingerprint {API} in {A}ndroid,'' in \emph{25th Annual
  Network and Distributed System Security Symposium, {NDSS'18}, San Diego,
  California, USA, February 18-21}, 2018.

\bibitem{zuodoes}
C.~Zuo, Z.~Lin, and Y.~Zhang, ``Why {D}oes {Y}our {D}ata {L}eak? {U}ncovering
  the {D}ata {L}eakage in {C}loud from {M}obile {A}pps,'' in \emph{IEEE
  Symposium on Security and Privacy, (SP'19), London, UK}, 2019.

\end{thebibliography}
%
\vspace{-12mm}
\begin{IEEEbiography}[{\includegraphics[width=1in,height=1.25in,clip,keepaspectratio]{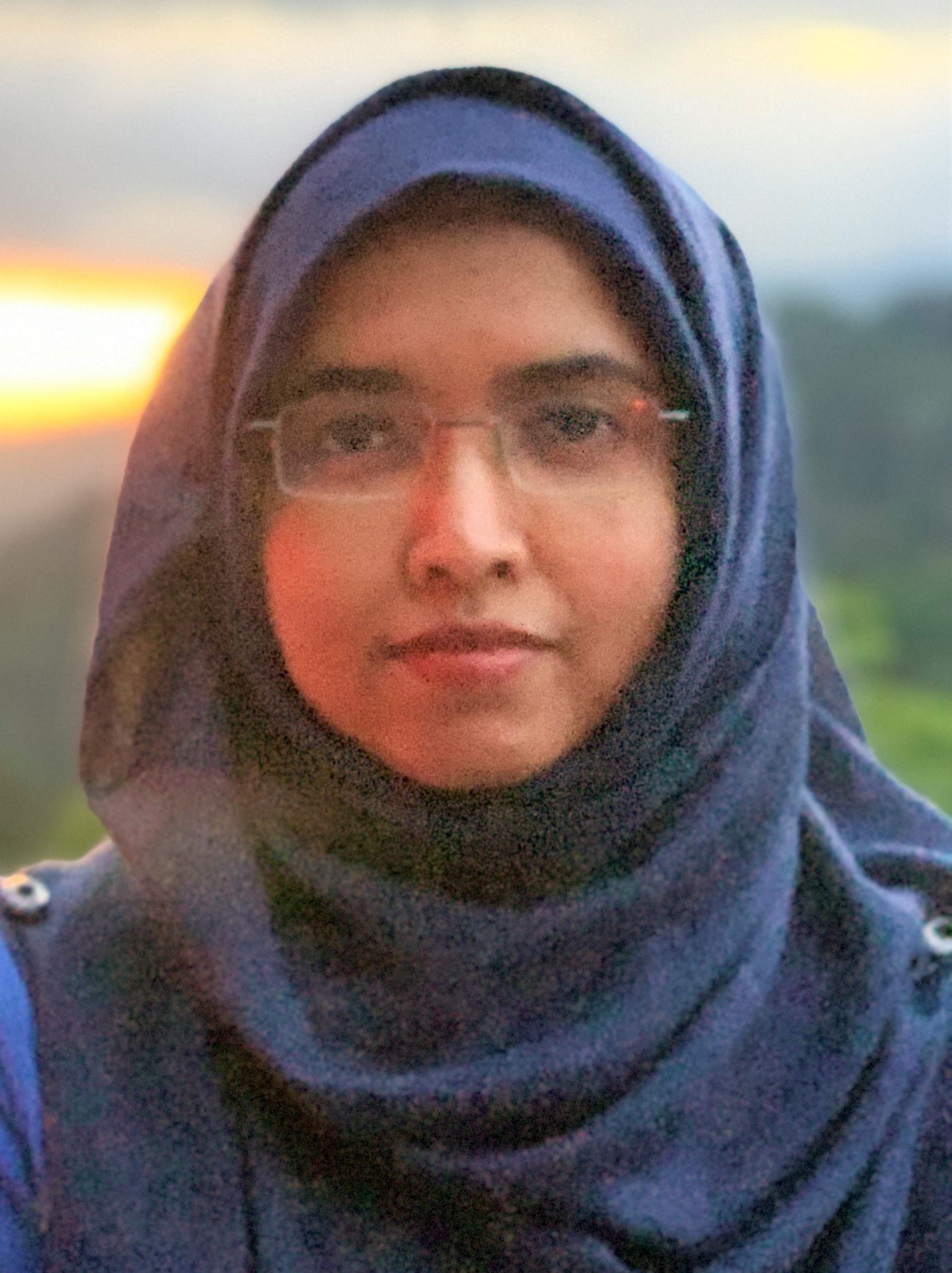}}]{Sharmin Afrose}
is a Ph.D. student in the department of computer science at Virginia Tech. She is working under the supervision of Professor Danfeng (Daphne) Yao. Her research interests include software security for Java Cryptographic API  and  AI bias in healthcare. She received a BS in Computer Science and Engineering from the Bangladesh University of Engineering and Technology (BUET). 
\end{IEEEbiography}

\vspace{-15mm}

\begin{IEEEbiography}[{\includegraphics[width=1in,height=1.25in,clip,keepaspectratio]{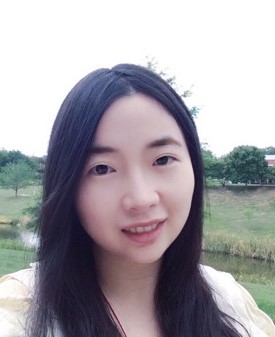}}]{Ya Xiao}
is  a  Ph.D.  student  of  computer  science department at Virginia Tech. She is working under the supervision of Professor Danfeng (Daphne) Yao. Her research interests include neural network based software security solutions, program analysis, and neural cryptanalysis. She has been awarded the Bit Shares Fellowship and the Dennis G. Kafura Fellowship at Virginia Tech.  She  received  M.S. degree  and  B.S. degree from Beijing University of Posts and Telecommunications (BUPT). 
\end{IEEEbiography}
\vspace{-14mm}
\begin{IEEEbiography}[{\includegraphics[width=1in,height=1.25in,clip,keepaspectratio]{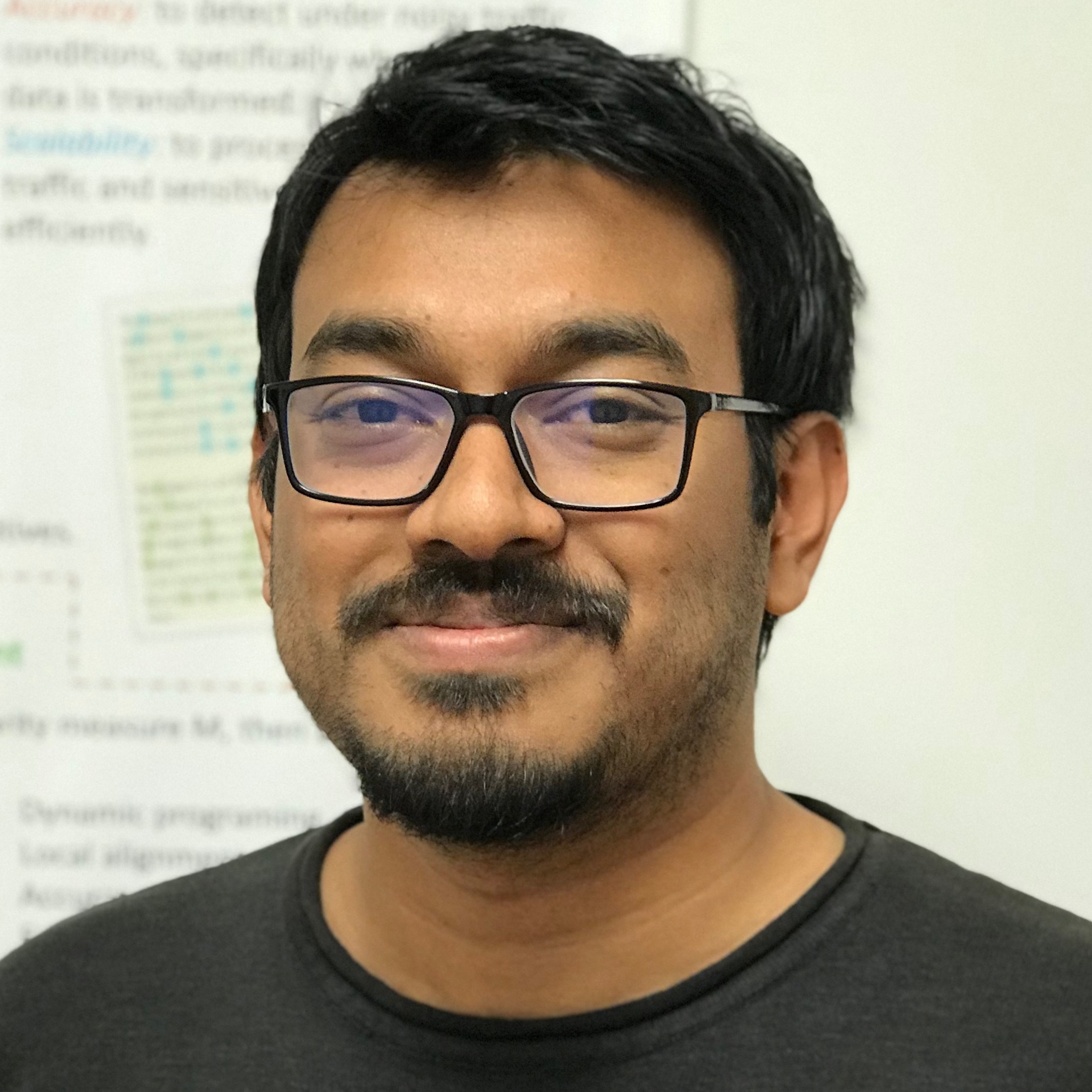}}]{Sazzadur Rahaman}
 is an assistant professor in the Department of Computer Science at the University of Arizona. He works towards making security research more democratized and affordable. He is broadly interested in computer security and privacy problems, specifically in building robust systems and methodologies by using program analysis, formal verification, applied cryptography, and machine learning-based techniques. Sazzadur completed his Ph.D. from Virginia Tech. He received his B.Sc. in computer science at Bangladesh University of Engineering and Technology (BUET). 
\end{IEEEbiography}

\vspace{-15mm}
\begin{IEEEbiography}[{\includegraphics[width=1in,height=1.25in,clip,keepaspectratio]{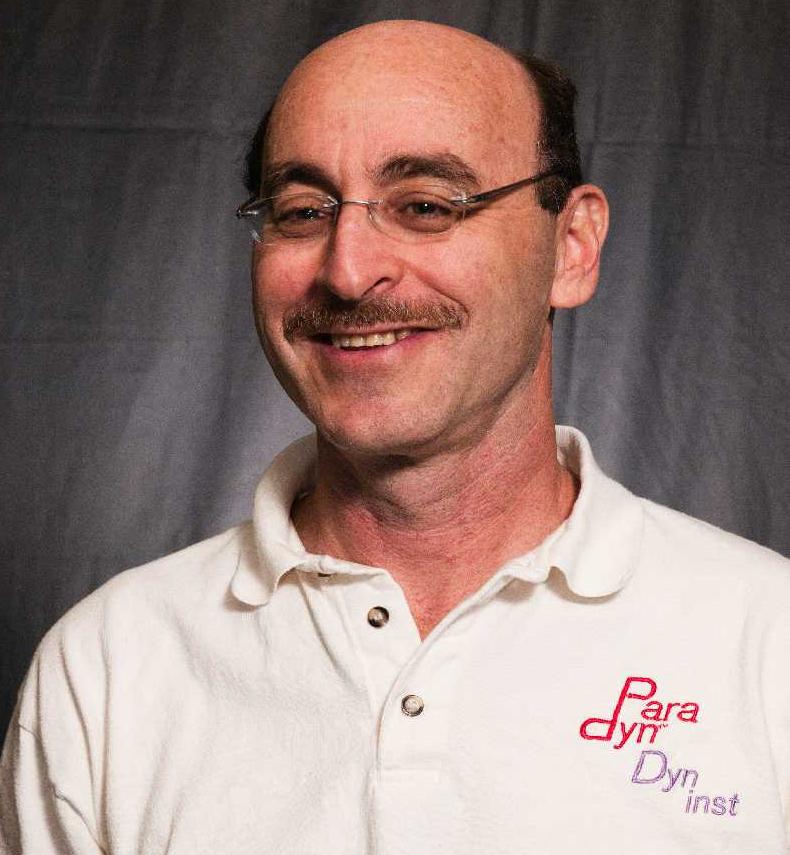}}]{Barton P. Miller} is a Vilas Distinguished Achievement Professor, and Amar \& Belinder Sohi Professor in Computer Sciences at the University of Wisconsin, Madison. He directs the Paradyn Tools project, which is investigating program scalability and binary program analysis and instrumentation technologies for use in HPC, systems design, and cyber-security. In addition, he is Chief Scientist of the DHS-funded Software Assurance Marketplace (SWAMP) research center. He also directs the vulnerability assessment and training program for the NSF Cybersecurity Center of Excellence, the Center for Trustworthy Scientific Cyberinfrastructure. He also co-directs the MIST software vulnerability assessment project in collaboration with his colleagues at the Autonomous University of Barcelona. He received his B.A. degree from the University of California, San Diego in 1977, and M.S. and Ph.D. degrees in Computer Science from the University of California, Berkeley in 1980 and 1984. Professor Miller is a Fellow of the ACM.
\end{IEEEbiography}
\vspace{-15mm}
\begin{IEEEbiography}[{\includegraphics[width=1in,height=1.25in,clip,keepaspectratio]{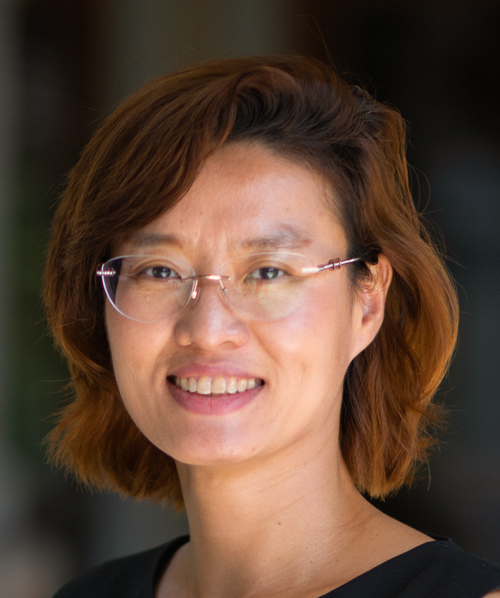}}]{Danfeng (Daphne) Yao} is a Professor of Computer Science at Virginia Tech. She is an Elizabeth and James E. Turner Jr. ’56 Faculty Fellow and CACI Faculty Fellow. Her research interests are on building deployable and proactive cyber defenses, focusing on detection accuracy and scalability. She has multiple US
patents for her inventions on network causal analysis for forensics. Dr. Yao received the NSF CAREER Award for her work on human-behavior driven malware detection and ARO Young Investigator Award for her semantic reasoning for mission-oriented security work. Dr. Yao is the ACM SIGSAC Treasurer/Secretary and is a member of the ACM SIGSAC
executive committee since 2017. She spearheads multiple inclusive excellence initiatives in the
cybersecurity research community, including the NSF-sponsored Individualized Cybersecurity
Research Mentoring (iMentor) Workshop and the Women in Cybersecurity Research (CyberW) Workshop. Daphne received her Ph.D. degree from Brown University, M.S. degrees from Princeton University and Indiana University, Bloomington, B.S. degree from Peking University
in China.
\end{IEEEbiography}







\end{document}